\documentclass{article}
\usepackage[utf8]{inputenc}
\usepackage{ifthen} 
\usepackage{subfigure}
\usepackage{longtable}
\usepackage[abs]{overpic}
\usepackage{textcomp}
\usepackage{pdflscape}
\usepackage[utf8]{inputenc}
\usepackage{fullpage}
\usepackage{graphicx}
\usepackage{tikz}
\usepackage{pgfplots}
\usepackage{booktabs}
\usepackage{amsmath}
\usepackage{multirow}
\usepackage{rotating}
\usepackage{appendix}
\usepackage{tabularx}
\newboolean{pdflatex}

\usepackage{jheppub}
\pdfoutput=1
\newcommand*{\GeV}{\ensuremath{\text{Ge\kern -0.1em V}}}

\title{Measurement of the muon flux for the SHiP experiment}
\collaborationImg{\includegraphics[width=2cm]{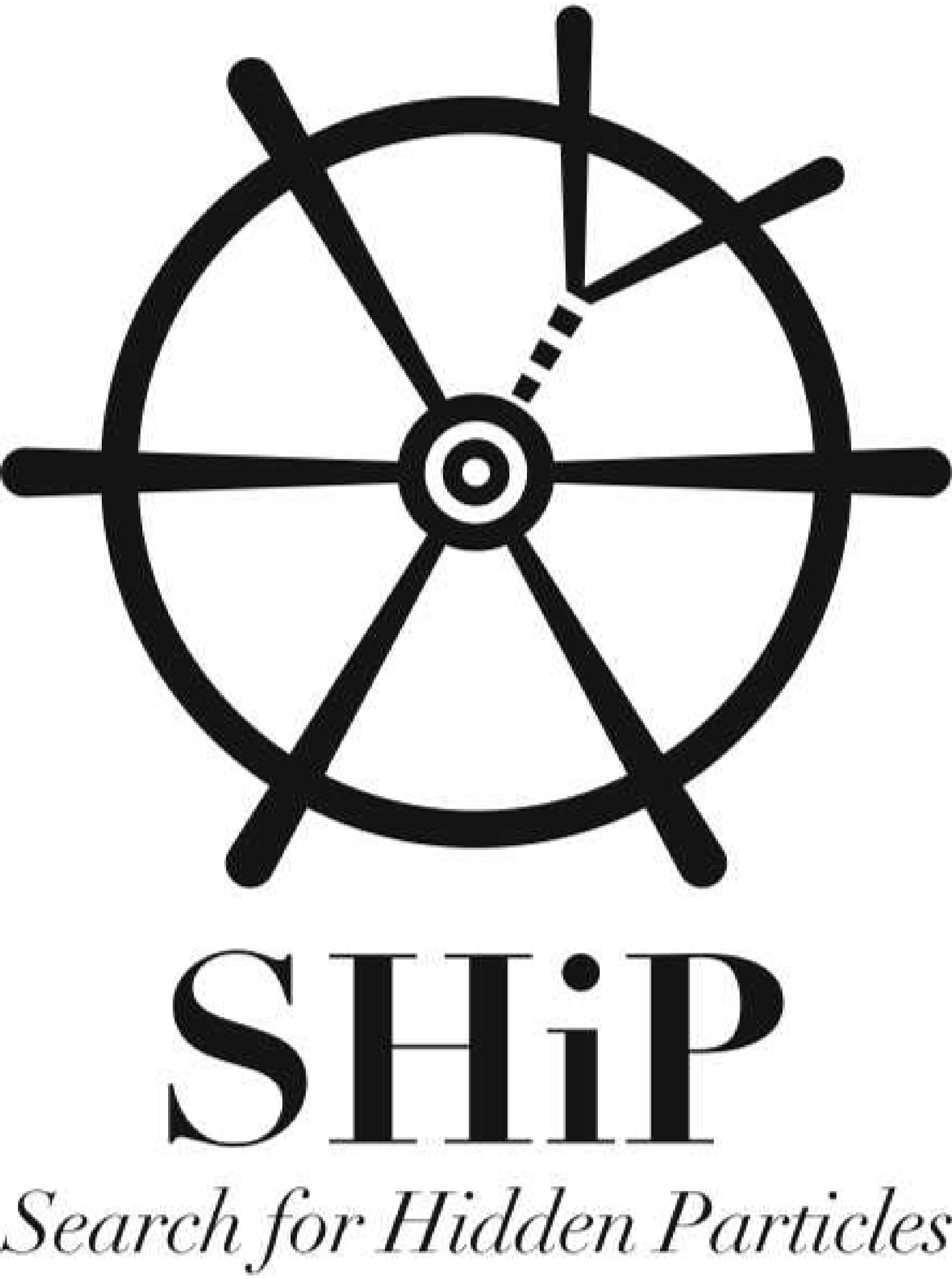}}
\collaboration{The SHiP Collaboration }
\emailAdd{eric.van.herwijnen@cern.ch}
\date{November 2019}
\author[44]{C.~Ahdida}
\author[48]{A.~Akmete}
\author[14,d,h]{R.~Albanese}
\author[14,32,34,d]{A.~Alexandrov}
\author[39]{A.~Anokhina}
\author[18]{S.~Aoki}
\author[44]{G.~Arduini}
\author[38]{E.~Atkin}
\author[29]{N.~Azorskiy}
\author[54]{J.J.~Back}

% ...
\affiliation[14]{Sezione INFN di Napoli, Napoli, Italy}
% ...
\affiliation[18]{Kobe University, Kobe, Japan}
% ...
\affiliation[29]{Joint Institute for Nuclear Research (JINR), Dubna, Russia}
% ...
\affiliation[32]{P.N.~Lebedev Physical Institute (LPI RAS), Moscow, Russia}
% ...
\affiliation[34]{National University of Science and Technology "MISiS", Moscow, Russia}
% ...
\affiliation[38]{National Research Nuclear University (MEPhI), Moscow, Russia}
\affiliation[39]{Skobeltsyn Institute of Nuclear Physics of Moscow State University (SINP MSU), Moscow, Russia}
% ...
\affiliation[44]{European Organization for Nuclear Research (CERN), Geneva, Switzerland}
% ...
\affiliation[48]{Middle East Technical University (METU), Ankara, Turkey}
% ...
\affiliation[54]{University of Warwick, Warwick, United Kingdom}
% ...
\affiliation[d]{Universit\`{a} di Napoli ``Federico II'', Napoli, Italy}
% ...
\affiliation[h]{Consorzio CREATE, Napoli, Italy}
% ...

\abstract{The SHiP experiment will search for very weakly interacting particles beyond the Standard Model which are produced in a 400 \GeV/$c$ proton beam dump at the CERN SPS. About $10^{11}$ muons per spill will be produced in the dump. To design the experiment such that the muon-induced background is minimized, a precise knowledge of the muon spectrum is required. To validate the muon flux generated by our Pythia and GEANT4 based Monte Carlo simulation (FairShip), we have measured the muon flux emanating from a SHiP-like target at the SPS.
This target, consisting of 13 interaction lengths of slabs of molybdenum and tungsten, followed by a 2.4 m iron hadron absorber was placed in the H4 400~\GeV/$c$ proton beam line. To identify muons and to measure the momentum spectrum, a spectrometer instrumented with drift tubes and a muon tagger were used.  During a three-week period a dataset for analysis corresponding to $(3.27\pm0.07)~\times~10^{11}$ protons on target was recorded. This amounts to approximatively 1\% of a SHiP spill. 
\\
\\
\textcopyright~CERN for the benefit of the SHiP collaboration.
}

\notoc

\pgfplotsset{compat=1.13}
\usepackage{ifthen}
\usepackage{microtype}
\usepackage{lineno}  
\usepackage{xspace} 
\usepackage{tabularx}
\usepackage{graphicx}  
\usepackage{color}
\usepackage{colortbl}
\graphicspath{{./figs/}} 

\usepackage{subfigure}

\usepackage{amsmath} 
\usepackage{amssymb}
\usepackage{amsfonts}
\usepackage{upgreek} 

\newcommand*\patchAmsMathEnvironmentForLineno[1]{%
\expandafter\let\csname old#1\expandafter\endcsname\csname #1\endcsname
\expandafter\let\csname oldend#1\expandafter\endcsname\csname
end#1\endcsname
 \renewenvironment{#1}%
   {\linenomath\csname old#1\endcsname}%
   {\csname oldend#1\endcsname\endlinenomath}%
}
\newcommand*\patchBothAmsMathEnvironmentsForLineno[1]{%
  \patchAmsMathEnvironmentForLineno{#1}%
  \patchAmsMathEnvironmentForLineno{#1*}%
}
\AtBeginDocument{%
\patchBothAmsMathEnvironmentsForLineno{equation}%
\patchBothAmsMathEnvironmentsForLineno{align}%
\patchBothAmsMathEnvironmentsForLineno{flalign}%
\patchBothAmsMathEnvironmentsForLineno{alignat}%
\patchBothAmsMathEnvironmentsForLineno{gather}%
\patchBothAmsMathEnvironmentsForLineno{multline}%
}

\usepackage{hyperref}    
\usepackage[all]{hypcap} 

\usepackage{relsize}
\usepackage{lipsum}

\def\ifmonospace{\ifdim\fontdimen3\font=0pt }

\def\C++{%
\ifmonospace%
    C++%
\else%
    C\kern-.1667em\raise.30ex\hbox{\smaller{++}}%
\fi%
\spacefactor1000 }

\def\Csharp{%
\ifmonospace%
    C\#%
\else%
    C\kern-.1667em\raise.30ex\hbox{\smaller{\#}}%
\fi%
\spacefactor1000 }

\begin{document}

\maketitle
\section{Introduction}
\label{sec:setup}
The aim of the SHiP experiment~\cite{TP} is to search for very weakly interacting particles beyond the Standard Model which are produced by the interaction of 400~\GeV/$c$ protons from the CERN SPS with a beam dump. The SPS will deliver $4\times10^{13}$ protons on target (POT) per spill, with the aim of accumulating $2\times10^{20}$ POT during five years of operation. The target is composed of a mixture of TZM (Titanium-Zirconium doped Molybdenum, $3.6 \lambda$\footnote{$\lambda$ is the interaction length.}), W ($9.2 \lambda$) and Ta ($0.5 \lambda$) to increase the
charm cross-section relative to the total cross-section and to reduce the
probability that long-lived hadrons decay.

An essential task for the experiment is to keep the Standard Model background level to less than 0.1 event after $2~\times~10^{20}$ POT. About $10^{11}$ muons per spill will be produced in the dump, mainly from the decay of $\pi, K, \rho, \omega$ and charmed mesons. These muons would give rise to a serious background for many hidden particle searches, and hence their flux has to be reduced as much as possible.  To achieve this, SHiP will employ a novel magnetic shielding concept~\cite{mushield} that will suppress the background by five orders of magnitude.
The design of this
shield relies on the precise knowledge of the kinematics of the produced muons, in particular the muons with a large momentum ($>$100~\GeV/$c$) and a large transverse momentum ($>$3~\GeV/$c$) as they can escape the
shield and end up in the
detector acceptance.

To validate the muon spectrum as predicted by our simulation, and hence the design of the shield, the SHiP Collaboration measured the muon flux in the experiment in the 400~\GeV/$c$ proton beam at the H4 beam line of the SPS at CERN in July 2018~\cite{EOI}.
\section{Experimental setup and data}
\subsection{Spectrometer}
\begin{figure}[ht]
\centering
\includegraphics[width=16cm]{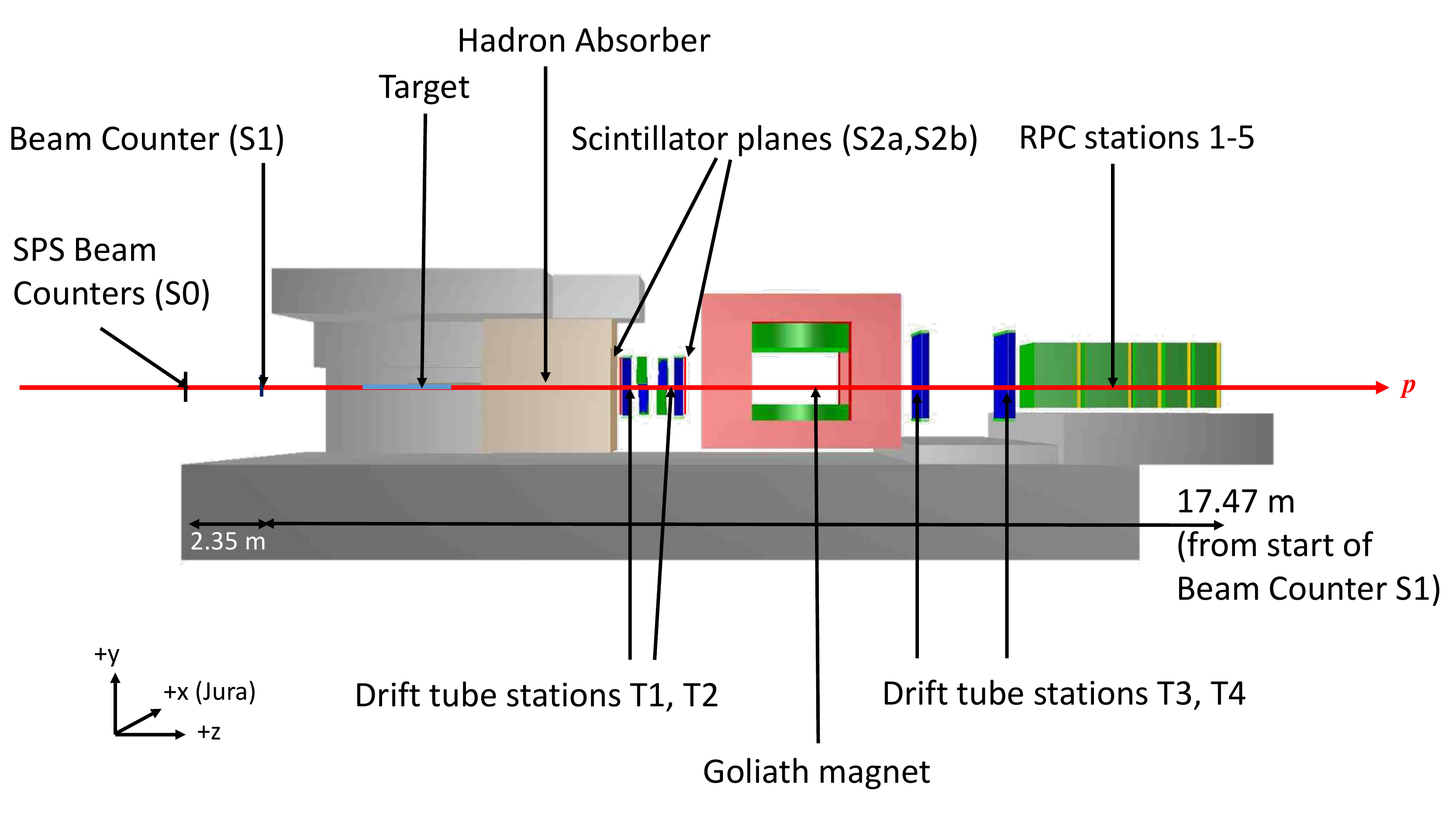}
\caption{Layout of the experimental setup to measure the $\mu$-flux. The FairShip (the SHiP software framwork) coordinate system is also shown.}
\label{fig:detector}
\end{figure}

The experimental setup, as implemented in FairShip (the SHiP software framework), is shown in Figure~\ref{fig:detector}.
A cylindrical SHiP-like\footnote{Without Ta cladding, but with thicker Mo and W slabs to preserve the same number of interaction lengths.} target  (10~cm diameter and 154.3~cm length) was followed by
a hadron absorber made of iron
blocks ($240 \times 240 \times 240~\rm{cm}^3$) and surrounded by iron and concrete shielding blocks. The dimensions of the hadron absorber were optimised to stop pions and kaons while keeping a good $p_T$ acceptance of traversing muons.
The SPS beam counters (XSCI.022.480/481, S0 in Figure~\ref{fig:detector}) and  beam counter S1 were used to count the number of POT seen by the experiment.

 A spectrometer was placed downstream of the hadron absorber. It consisted of four drift-tube stations (T1--T4, modified from the OPERA experiment \cite{opera}) with two stations upstream and two stations downstream of the Goliath magnet~\cite{goliath}. The drift-tubes were arranged in modules of 48 tubes, staggered in four layers of twelve tubes with a total width of approximately $50~\rm{cm}$. The four modules of height $110~\rm{cm}$ making up stations T1 and T2 were arranged in a stereo setup ($x-u$ views for T1 and $v-x$ views for T2), with a stereo angle of $60^{\circ}$. T3 and T4 had only $x$ views and were made of four modules of $160~\rm{cm}$ height.

The drift-tube trigger (S2) consisted of two scintillator planes, placed before (S2a) and behind (S2b) the first two tracking stations.

A muon tagger was placed behind the two downstream drift-tube stations. It consisted of five planes of single-gap resistive plate chambers (RPCs), operated in avalanche mode, interleaved with $1\times80~\rm{cm}$ and $3\times40~\rm{cm}$ thick iron slabs. In addition to this, a $80~\rm{cm}$ thick iron slab was positioned immediately upstream of the first chamber. The active area of the RPCs was $190~\rm{cm} \times 120~\rm{cm}$ and each chamber was read out by two panels of $x/y$ strips with a $1~\rm{cm}$ pitch.

The two upstream tracking stations were centered on the beam line, whereas the two downstream stations and the RPCs were centered on the Goliath magnet\footnote{The centre of the Goliath magnet is $17.86~\rm{cm}$ above the beam line.} opening to maximize the acceptance.

The data acquisition was triggered by the coincidence of S1 and S2. For more details on the DAQ framework, see~\cite{PG:DAQ}, and for a description of the trigger and the DAQ conditions during data taking, see~\cite{PG:Trigger}.

The protons were delivered in $4.8~\rm{s}$ duration spills (slow extraction). There were either one or two spills per SPS supercycle, with intensities $\sim3\times10^6$  protons per second.  The 1-sigma width of the beam spot was 2~mm. For physics analysis, 20128 useful  spills were recorded with the full magnetic field of 1.5~T, with $2.81\times10^{11}$ raw S1 counts. After normalization (see Section~\ref{sec:normalization}) this corresponds to $(3.25\pm0.07)\times10^{11}$ POT. Additional data were taken with the magnetic field switched off for detector alignment and tracking efficiency measurement.

\section{Data analysis}
\subsection{Normalization}
\label{sec:normalization}
The calculation of the number of POT delivered to the experiment must take
the different signal widths and dead times of the various scintillators into account. Moreover, some protons from the so-called halo, might fall outside the acceptance of S1 and will only be registered by S0.

In low-intensity runs these effects are small. We select some spills of these runs and split them into 50 slices of 0.1 s. We then determine the number of POT per slice and count the number of reconstructed muons in each slice, which should be independent of the intensity. By leaving the dead times as free parameters in a straight line fit, we find~\cite{norm} that the number of POT required to have an event with at least one reconstructed muon  is $710\pm15$. The systematic error of 15 POT accounts for the variation between the runs used for the normalization. The statistical error is negligible. The trigger inefficiency is less than 1\textperthousand~and is hence neglected. Multiplying the number of reconstructed muons found in the 20128 spills by 710 we calculated that this data set corresponds to $(3.25\pm0.07)~\times~10^{11}$ POT.

\subsection{Tracking}
\begin{figure}[ht]
\centering
\includegraphics[width=16cm]{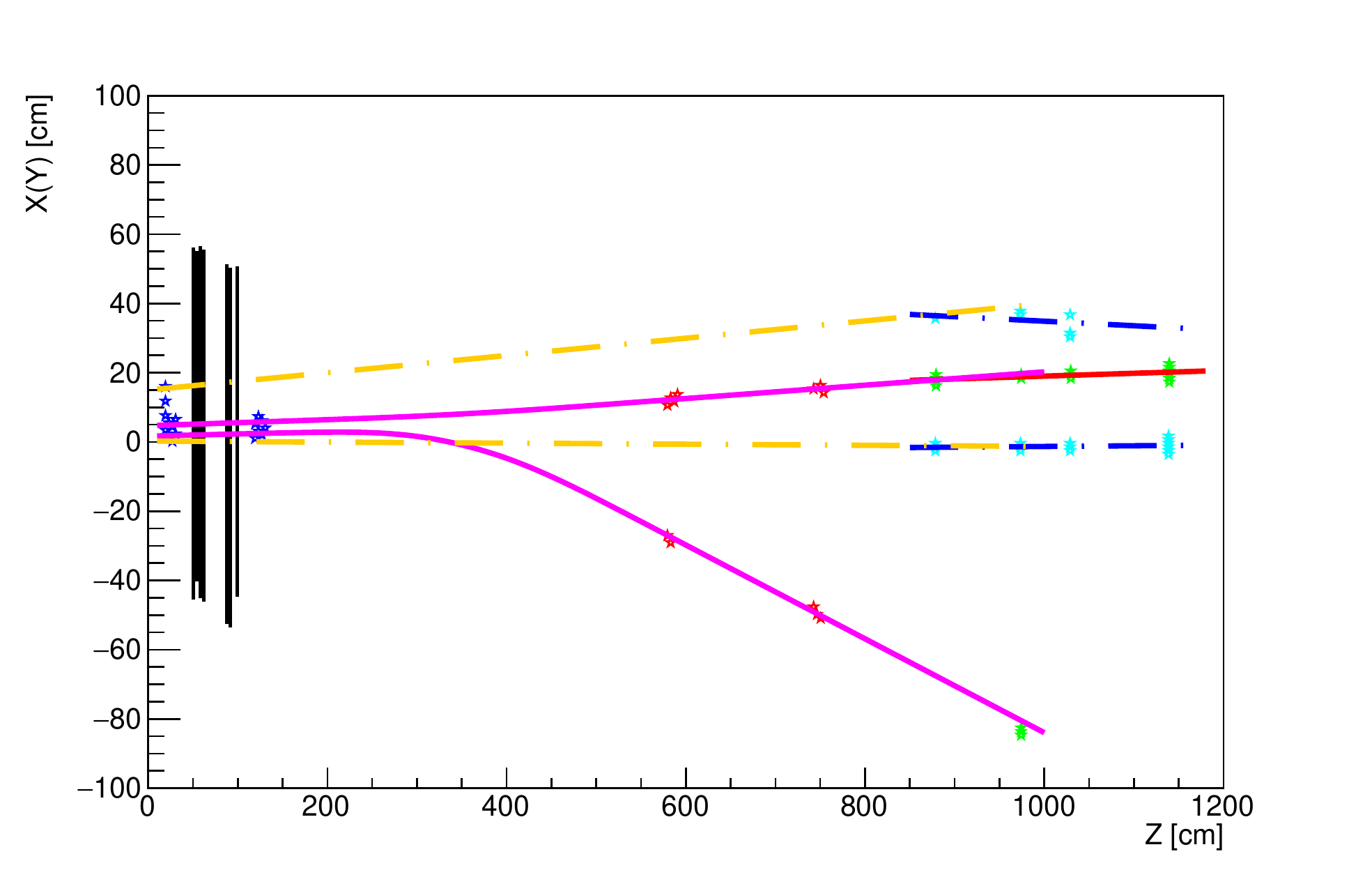}
\caption{A two-muon event in the event display. The blue crosses are hits in Drift-tube stations T1 and T2, the red crosses are hits in T3 and T4. The green and light blue are hits in the RPC stations. The orange (blue) dotted lines are drift tube (RPC) track segments in the $y$ projection; the pink (red) curves are track segments in the $x$ projection.}
\label{fig:eventdisplay}
\end{figure}

For the drift-tubes, the relation between the measured drift-time and the distance of the track to the wire (the "$r\text{-}t$" relation) is obtained from the Time to Digital Converter (TDC) distribution by assuming a uniformly illuminated tube. 
When reconstructing the data, the  $r\text{-}t$ relations are established first by looking the TDC distributions of simple events (i.e. events with at least 2 and a maximum of 6 hits per tracking station). In the simulation, the true drift radius is smeared with the expected resolution. The pattern recognition subsequently selects hits and clusters to form track candidates and provides the starting values for the track fit. The RPC pattern recognition proceeds similarly. drift-tube tracks are then extrapolated to RPC tracks and tagged as muons if they have hits in at least three RPC stations.
Figure~\ref{fig:eventdisplay} shows a two-muon event in the event display.

\subsection{Momentum resolution}
\begin{figure}[ht]
\centering
\includegraphics[width=9cm]{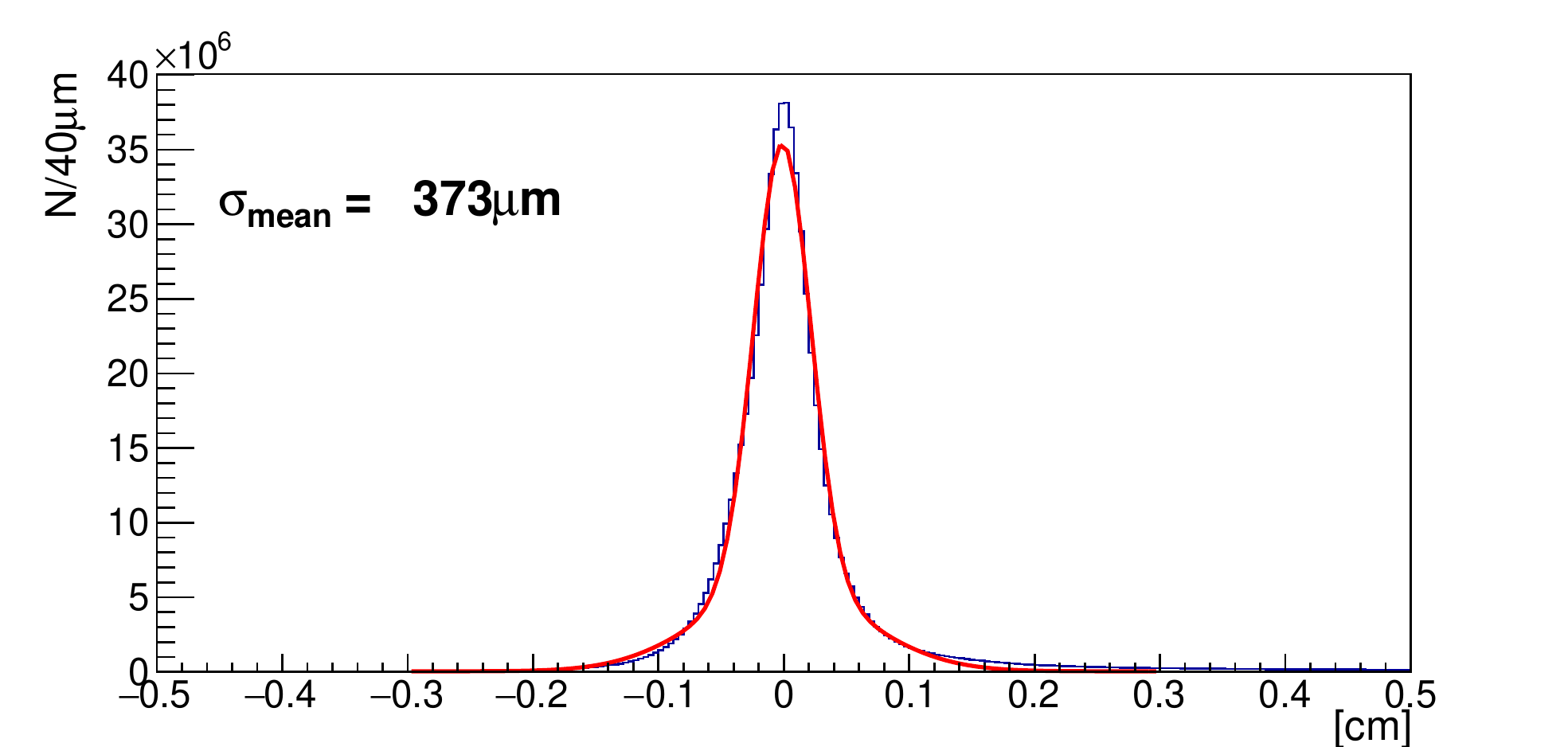}
\caption{Average of all drift-tube residuals. The fit is a double Gaussian and the resulting hit resolution ($\sigma_{\rm{mean}}$) is the average of the two sigma's.}
\label{fig:residuals}
\end{figure}

\begin{figure}[ht]
\centering
\label{fig:mcsmear}
\noindent\makebox[\textwidth]{
\includegraphics[width=9cm]{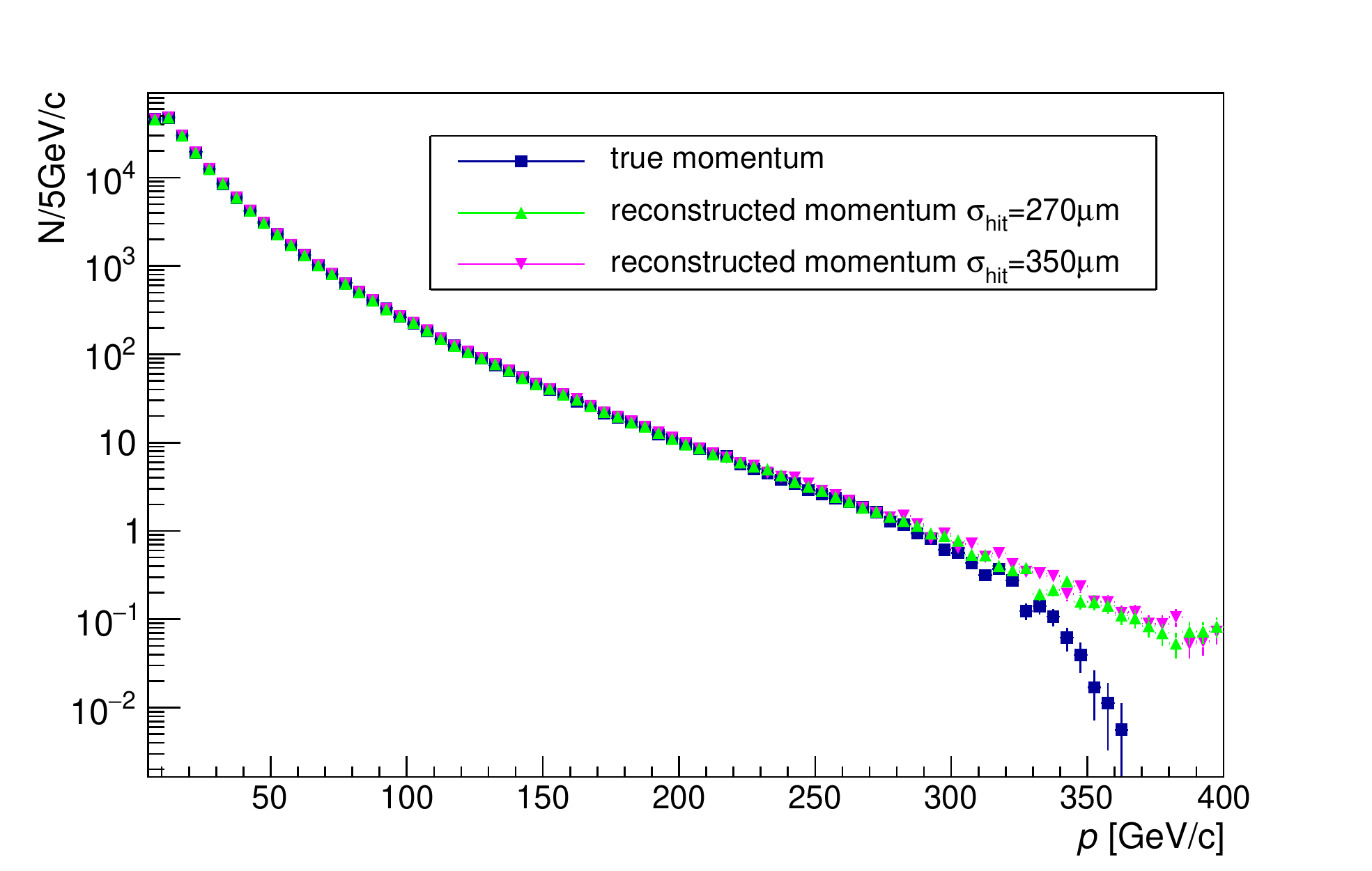}
\includegraphics[width=9cm]{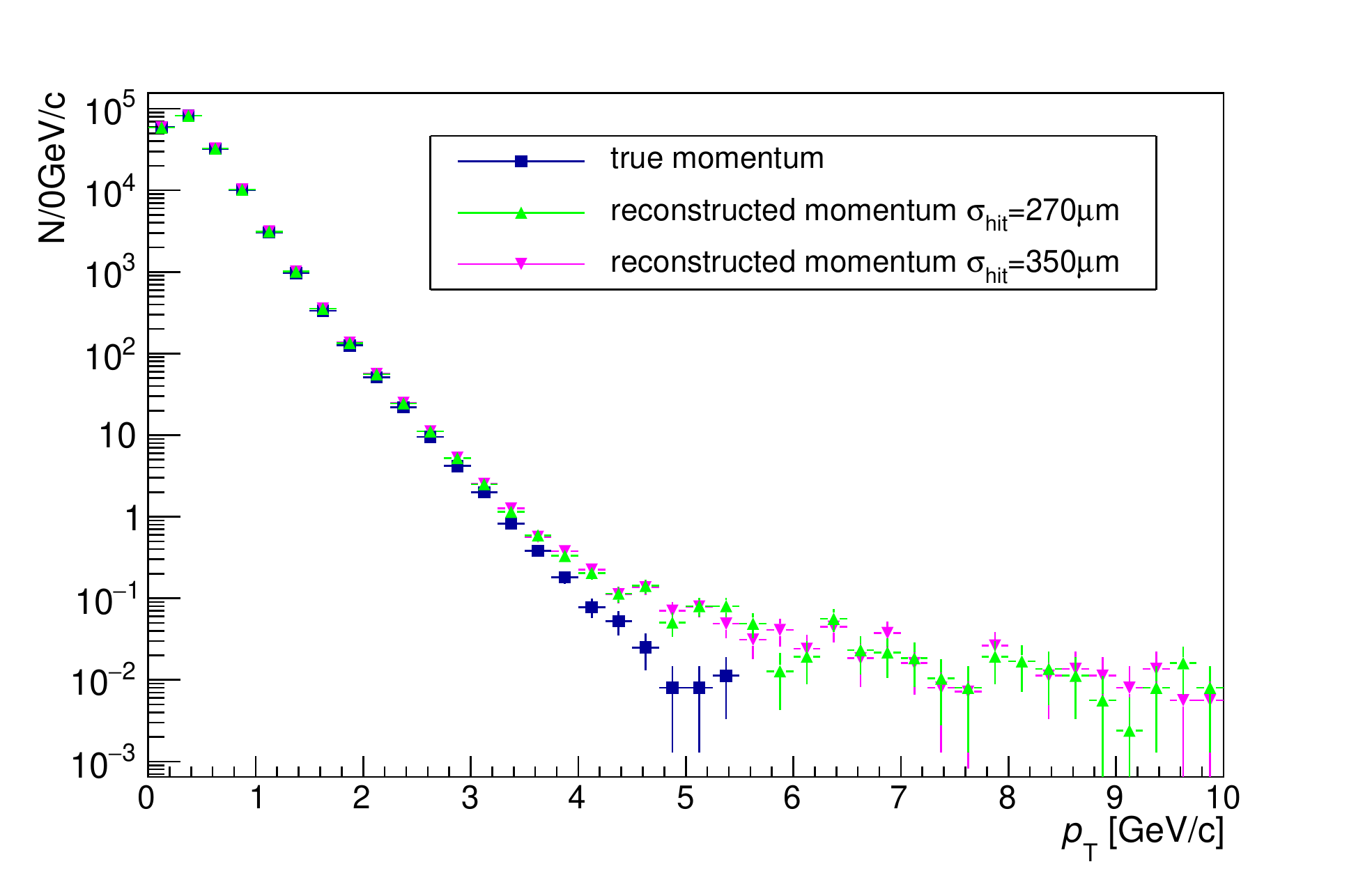}
}
\caption{Effect of additional Gaussian smearing on the momentum distribution in the simulation, left $p$, right $p_{T}$. The distributions correspond to the simulation truth before reconstruction (navy blue), the nominal resolution $\sigma_{\rm{hit}}=270~\mu \rm{m}$ (green) and a degraded resolution  $\sigma_{\rm{hit}}=350~\mu \rm{m}$ (pink).}
\end{figure}

The expected drift-tube hit resolution based on the OPERA results is 270~\textmu m~\cite{opera}. However, due to residual misalignment and imperfect $r\text{-}t$ relations, the measured hit resolution was slightly worse, 373~\textmu m, as shown in Figure~\ref{fig:residuals}.
To study the impact of degraded spatial drift-tube resolution the  momentum distribution from the simulation was folded with additional smearing as shown in Figure~\ref{fig:mcsmear}.
The tails towards large momentum $p$ and $p_{T}$ are caused mainly by tracks fitted with wrong drift times due to background hits.

From Figure~\ref{fig:mcsmear} we conclude that the momentum resolution is not strongly affected by the degraded resolution of the drift-tubes that is observed. The effect of the degraded drift-tube resolution is therefore negligible for our studies of the momentum spectrum. To account for residual effects in the track reconstruction, the resolution in the simulation was set to  350~\textmu m.

\subsection{Tracking efficiencies}
The tracking efficiency in the simulation depends on the station occupancy, and in data and simulation the occupancies are different (apparently caused by different amounts of delta rays). By taking this into account, the efficiency in the simulation is reduced from 96.6\% to 94.8\%.

To determine the tracking efficiency in data, we use the RPCs to identify muon tracks in the data with the magnetic field turned off. We then take the difference between the tracking efficiency in the simulation with magnetic field off (96.9\%) and the measured efficiency (93.6\%) as the systematic error: 3.3\%. For more details on the analysis and reconstruction, see~\cite{anal}.

\section{Comparison with the simulation}
A large sample of muons was generated (with Pythia6,  Pythia8~\cite{Pythia} and GEANT4~\cite{geant4} in FairShip) for the background studies of SHiP, corresponding to the number of POT as shown in Table~\ref{tab:MCPOT}. The energy cuts ($E_{\mathrm{min}}$) of 1~\GeV~and~10~\GeV~were imposed to save computing time. The primary proton nucleon interactions are simulated by Pythia8.
The emerging particles are transported by GEANT4 through the target and hadron absorber producing a dataset of also referred to as "mbias" events. A special setting of GEANT4 was used to switch on muon interactions to produce rare dimuon decays of low-mass resonances.
Since GEANT4 does not have production of heavy flavour in particle interactions, an extra procedure was devised to simulate heavy-flavour production not only in the primary $pN$ collision but also in collisions of secondary particles with the target nucleons. For performance reasons, this was done with Pythia6.  The mbias and charm/beauty datasets were combined by removing the heavy-flavour contribution from the mbias and inserting the cascade data with appropriate weights.  The details of the full heavy-flavour production for both the primary and cascade interactions are described in~\cite{cascade}.

\begin{table}[ht]
\centering
\caption{Simulation samples made for SHiP background studies. $\chi$ is the fraction of protons that produce heavy flavour.}
\label{tab:MCPOT}
\begin{tabular}{|l|l|r|}
\hline
$E_{\mathrm{kin}} > E_{\mathrm{min}}$ & mbias/cascade  &POT       \\
\hline 
\hline	
1~\GeV& mbias & $1.8~\times~10^9$ \\
1~\GeV & charm ($\chi_{c\overline{c}}=1.7~\times~10^{-3}$) & $10.2~\times~10^9$ \\
10~\GeV & mbias & $65.0~\times~10^9$ \\
10~\GeV & charm ($\chi_{c\overline{c}}=1.7~\times~10^{-3}$)& $153.3~\times~10^9$ \\
10~\GeV & beauty ($\chi_{b\overline{b}}=1.3~\times~10^{-7}$)& $5336.0~\times~10^9$ \\
\hline
\end{tabular}
\end{table}

\section{Results}
The main objective of this study is to validate our simulations for the muon background estimation for the SHiP experiment.
For this purpose, we compare the reconstructed momentum distributions ($p$ and $p_{T}$) from data and simulation.

As discussed in the previous section (see also Figure~\ref{fig:mcsmear}), the events outside the limits ($p>350~$\GeV/$c$ or $p_{T}>5~$\GeV/$c$) are dominated by wrongly reconstructed trajectories due to background hits and the limited precision of the tracking detector. In SHiP, where the hadron absorber is 5~m long, only muons with momentum $p>5~$\GeV/$c$  have sufficient energy to traverse the entire absorber. We therefore restrict our comparison to 5~\GeV/$c<p<300~$\GeV/$c$ and $p_{T}<4~$\GeV/$c$. For momenta below 10~\GeV/$c$, we only rely on the reconstruction with the tracking detector, since they do not reach the RPC stations. Above 10~\GeV/$c$ we require the matching between drift-tube and RPC tracks.

\begin{figure}[ht]
\centering
\includegraphics[width=\textwidth]{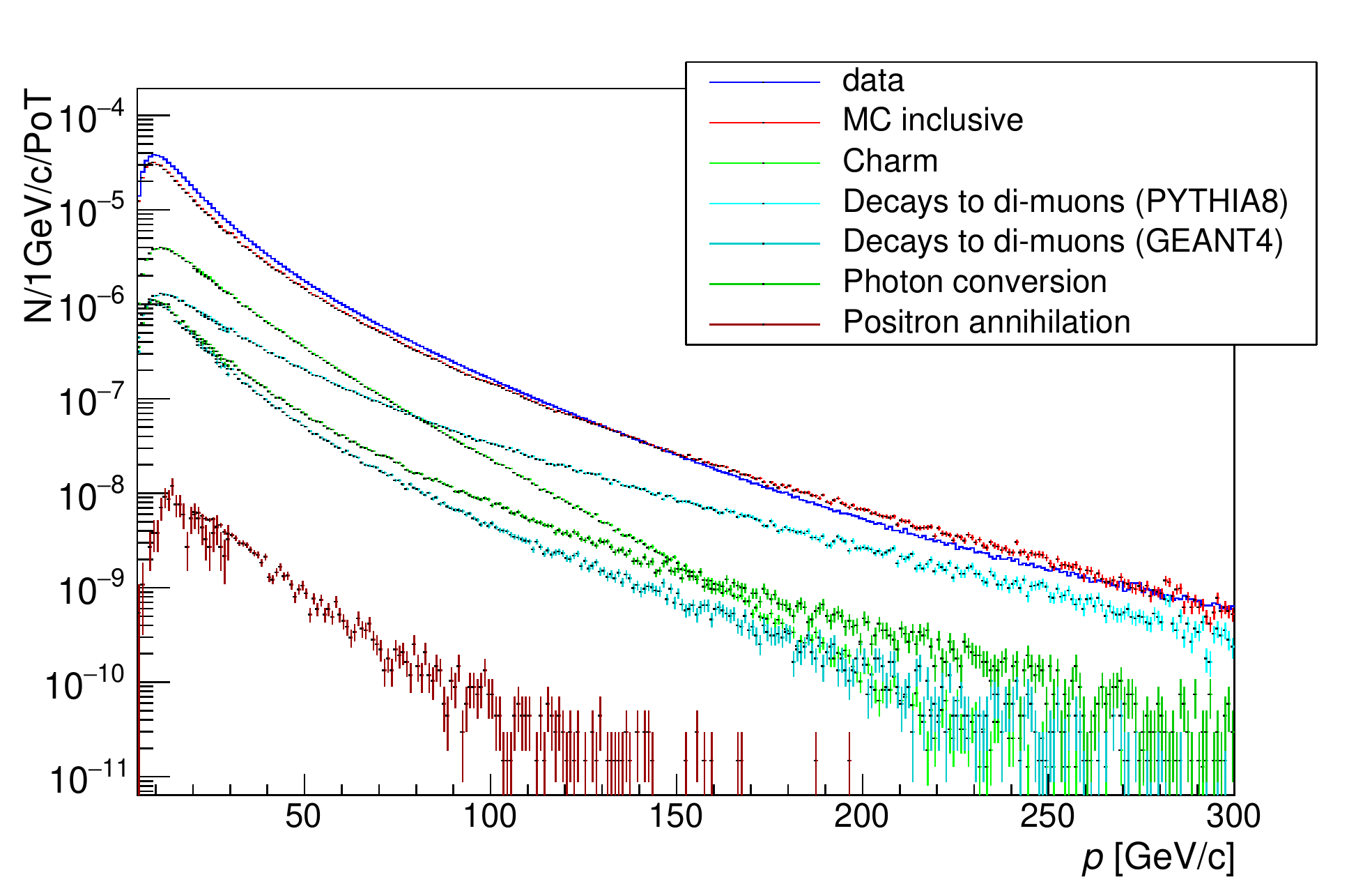}
\includegraphics[width=\textwidth]{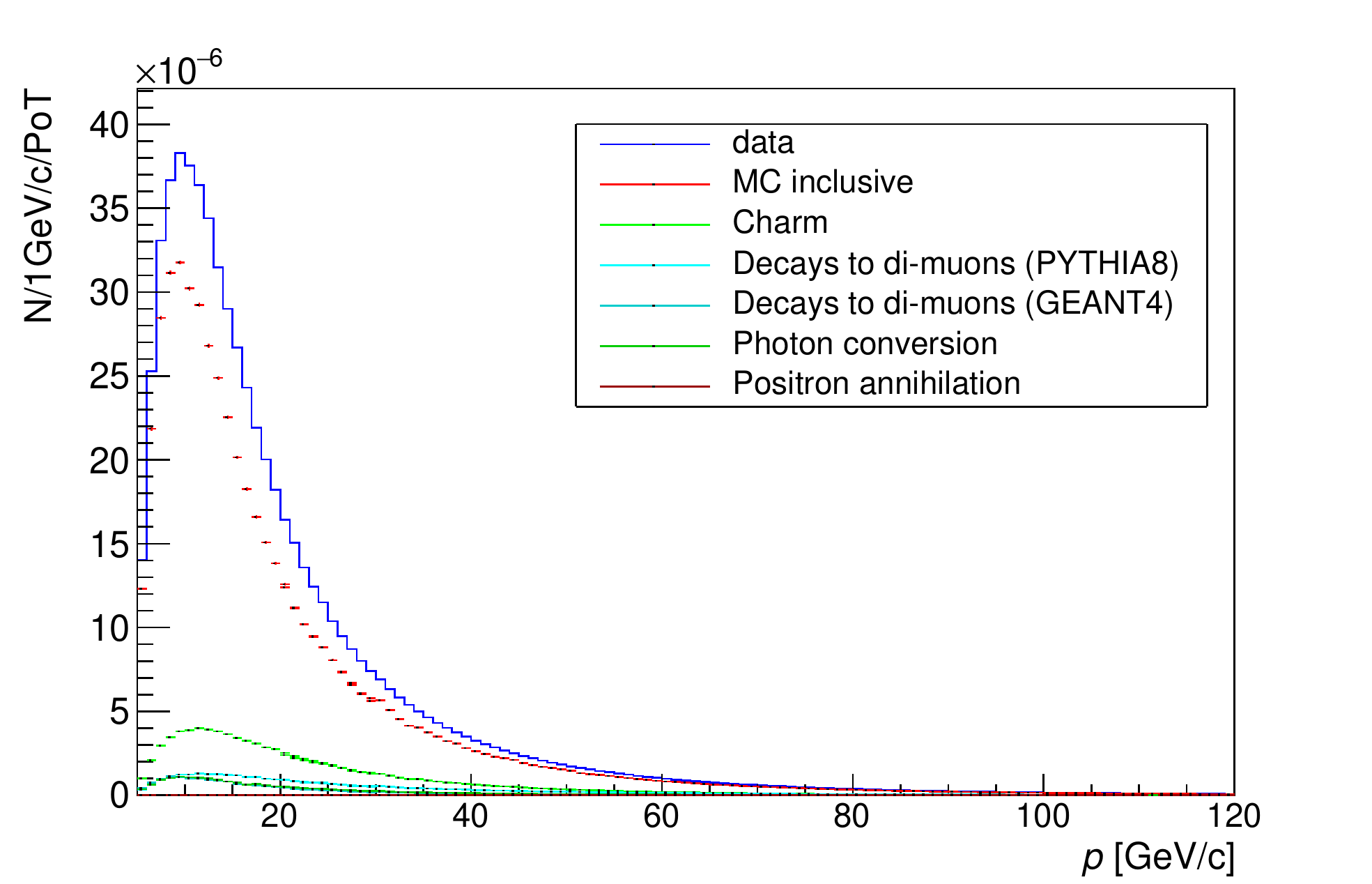}
\caption{Measured muon momentum distributions from data and simulation, top full range in log scale, bottom detail of the low momentum range with a linear scale. The distributions are normalized to the number of POT. For simulated data, some individual sources are highlighted, muons from charm (green), from dimuon decays of low-mass resonances in Pythia8 (cyan), in Geant4 (turquoise), photon conversion (dark green) and positron annihilation (brown).}
\label{fig:momplotsP}
\end{figure}

\begin{figure}[ht]
\centering
\includegraphics[width=\textwidth]{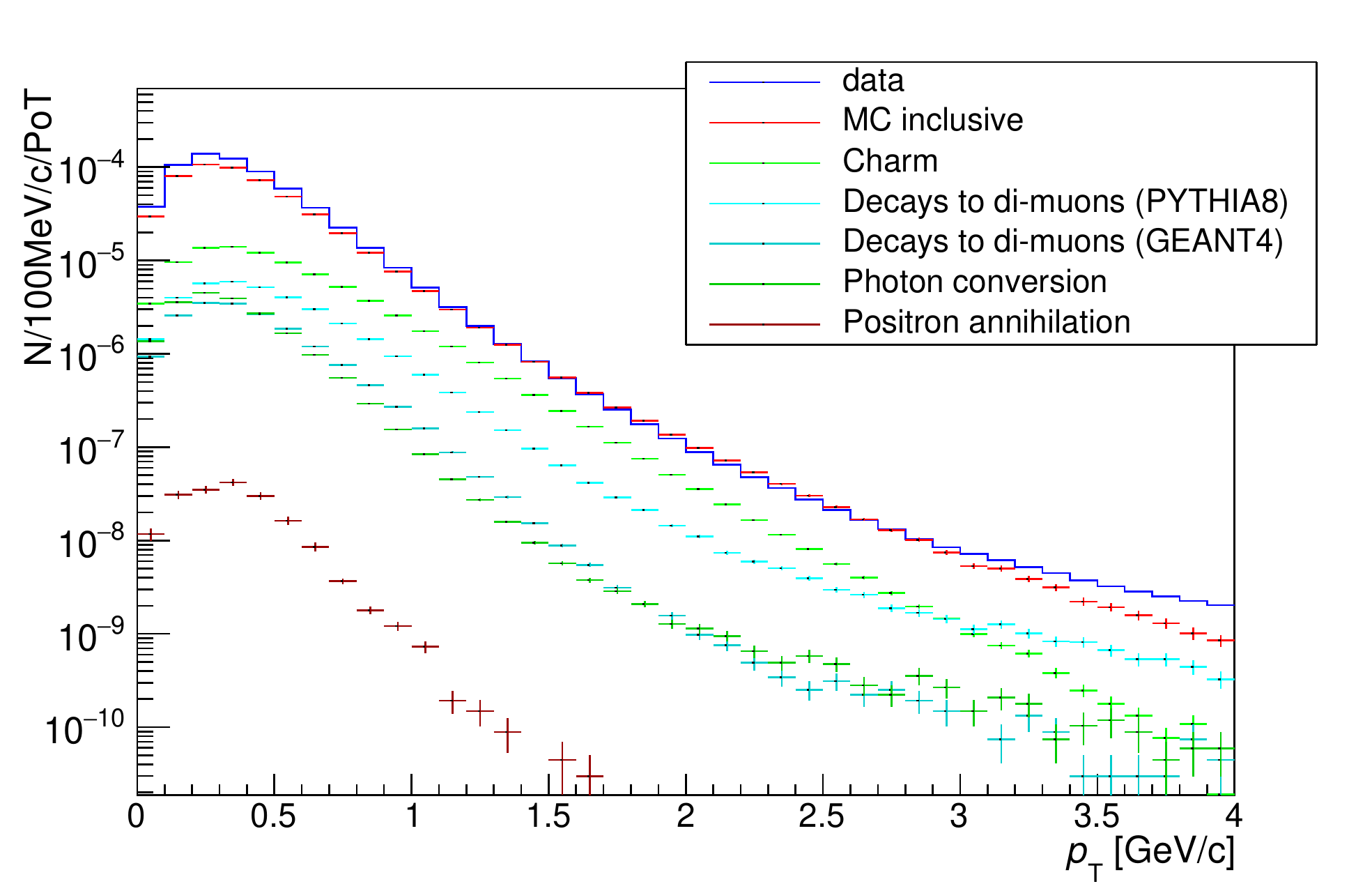}
\includegraphics[width=\textwidth]{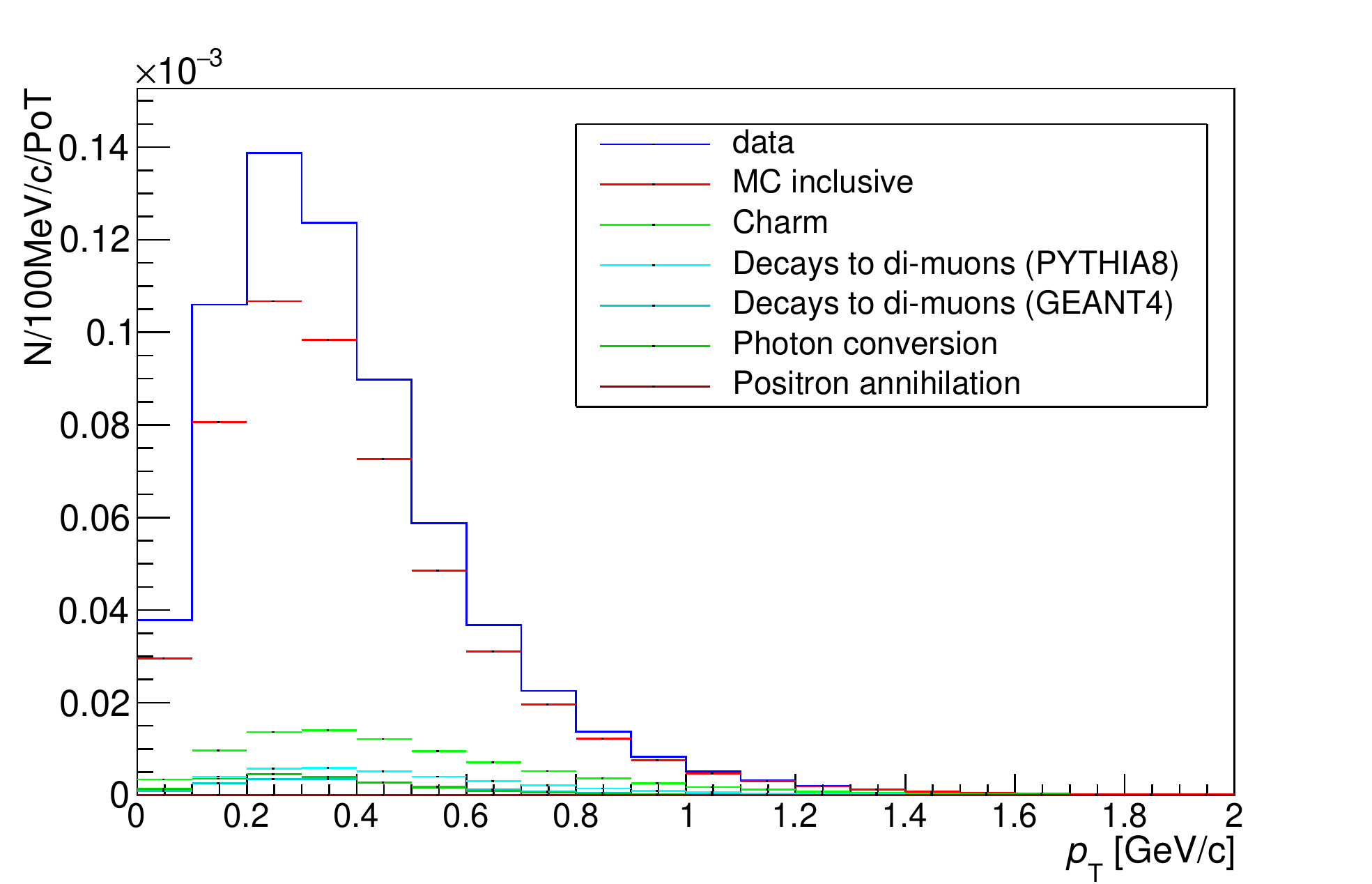}
\caption{Transverse momentum distributions from data and simulation, top full range in log scale, bottom detail of lower transverse momentum with a linear scale. The distributions are normalized to the number of POT. For the simulation, some individual sources are highlighted, muons from charm (green), from dimuon decays of low-mass resonances in Pythia8 (cyan), in Geant4 (turquoise), photon conversion (dark green) and positron annihilation (brown). }
\label{fig:momplotsPt}
\end{figure}

Figure~\ref{fig:momplotsP} and Figure~\ref{fig:momplotsPt} show the $p$ and $p_{T}$ distributions of muon tracks. The distributions are normalized to the number of POT for data (see Section~\ref{sec:normalization}) and simulation respectively. For the simulated sample, muons from some individual sources are also shown in addition to their sum.

\begin{figure}[ht]
\centering
\includegraphics[width=0.4\textwidth]{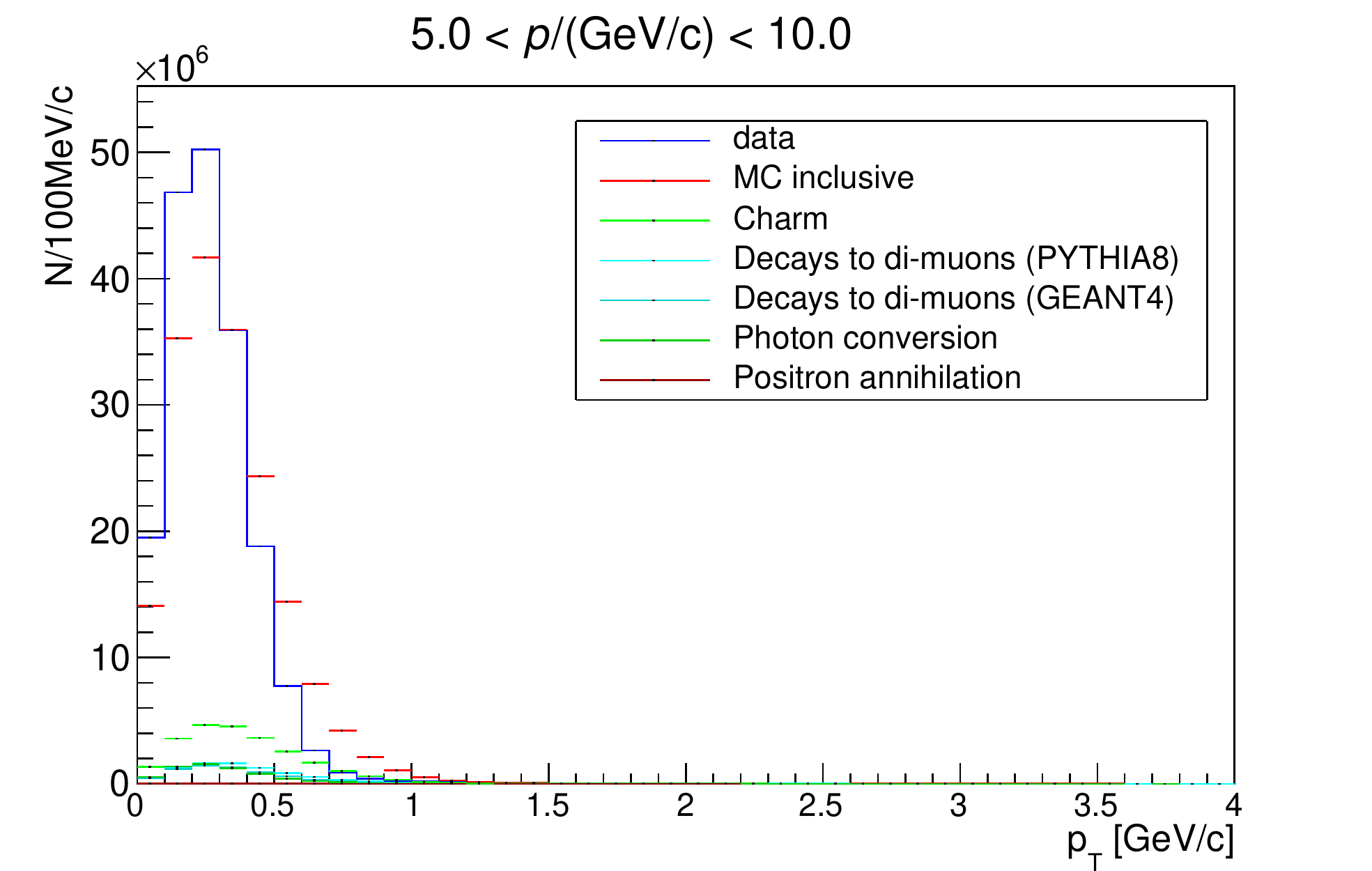}\includegraphics[width=0.4\textwidth]{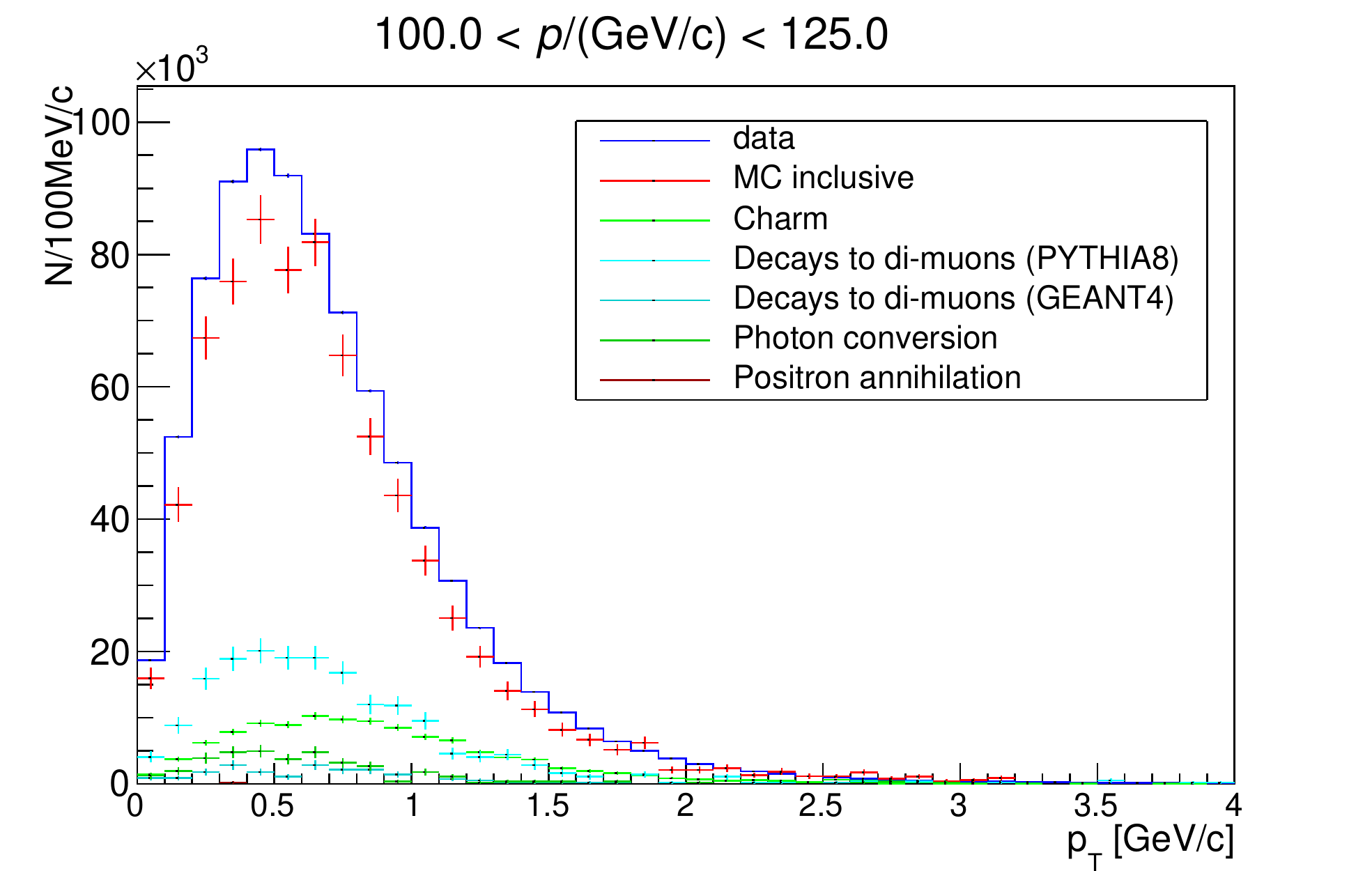}

\includegraphics[width=0.4\textwidth]{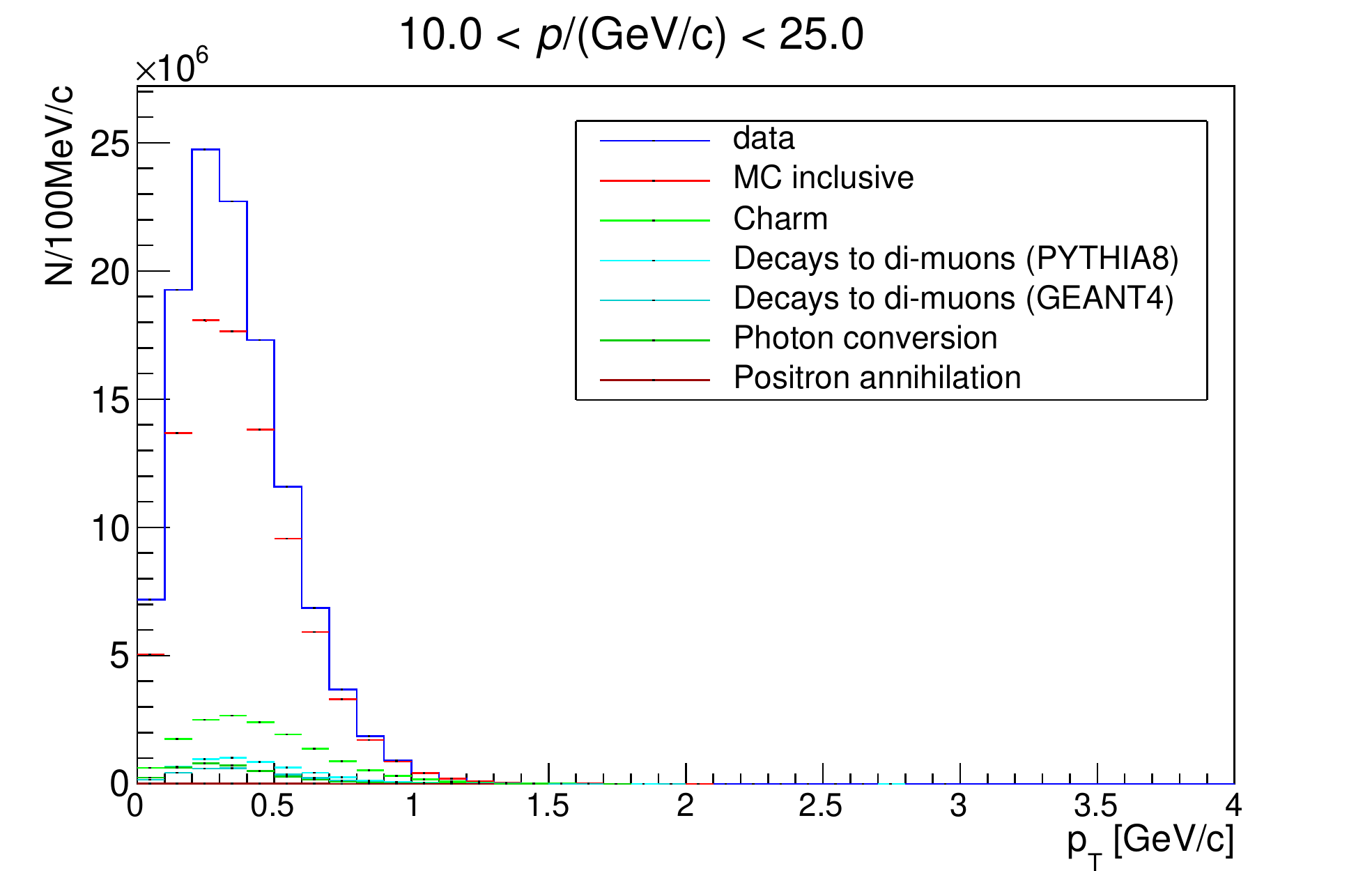}\includegraphics[width=0.4\textwidth]{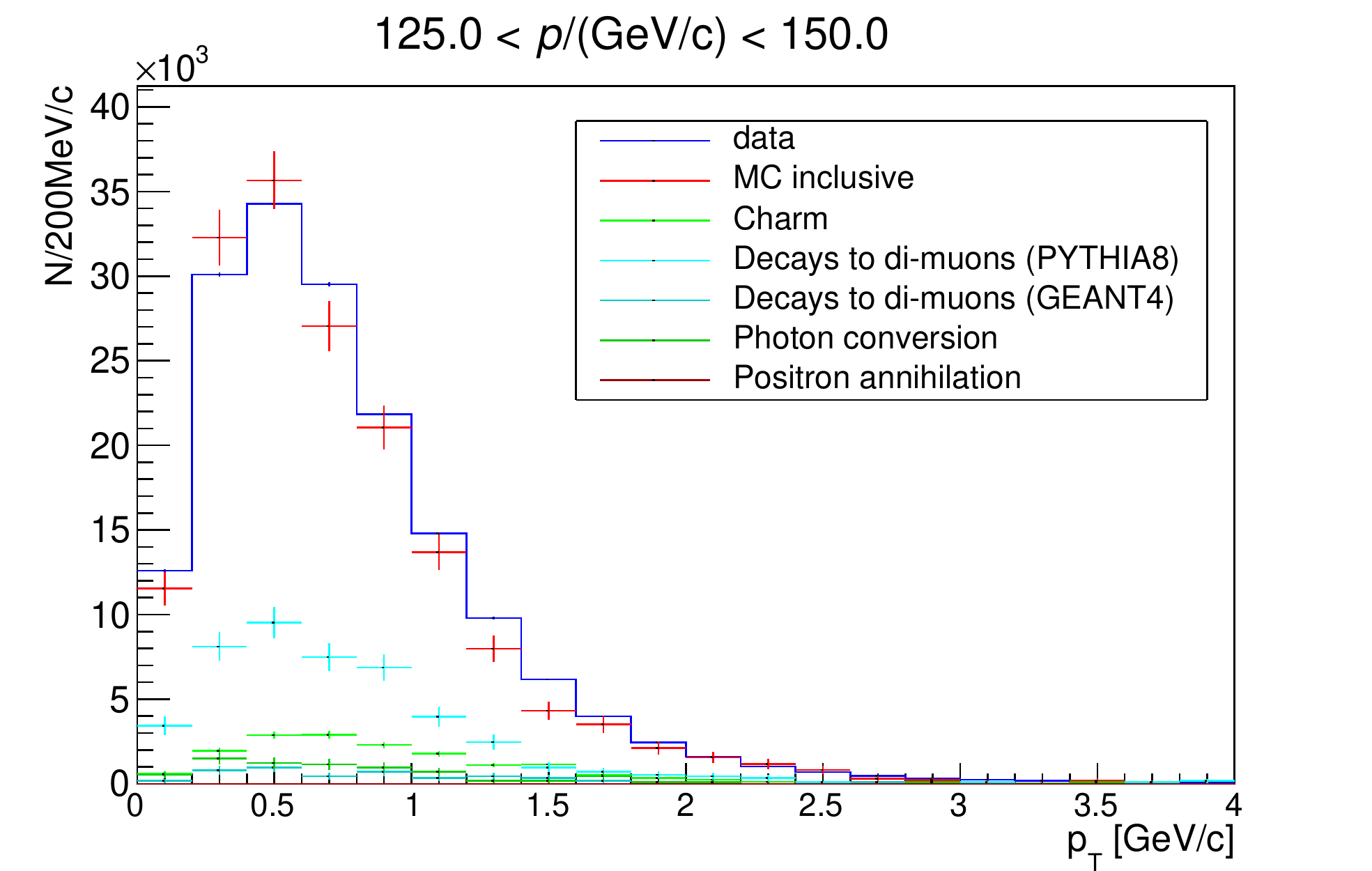}

\includegraphics[width=0.4\textwidth]{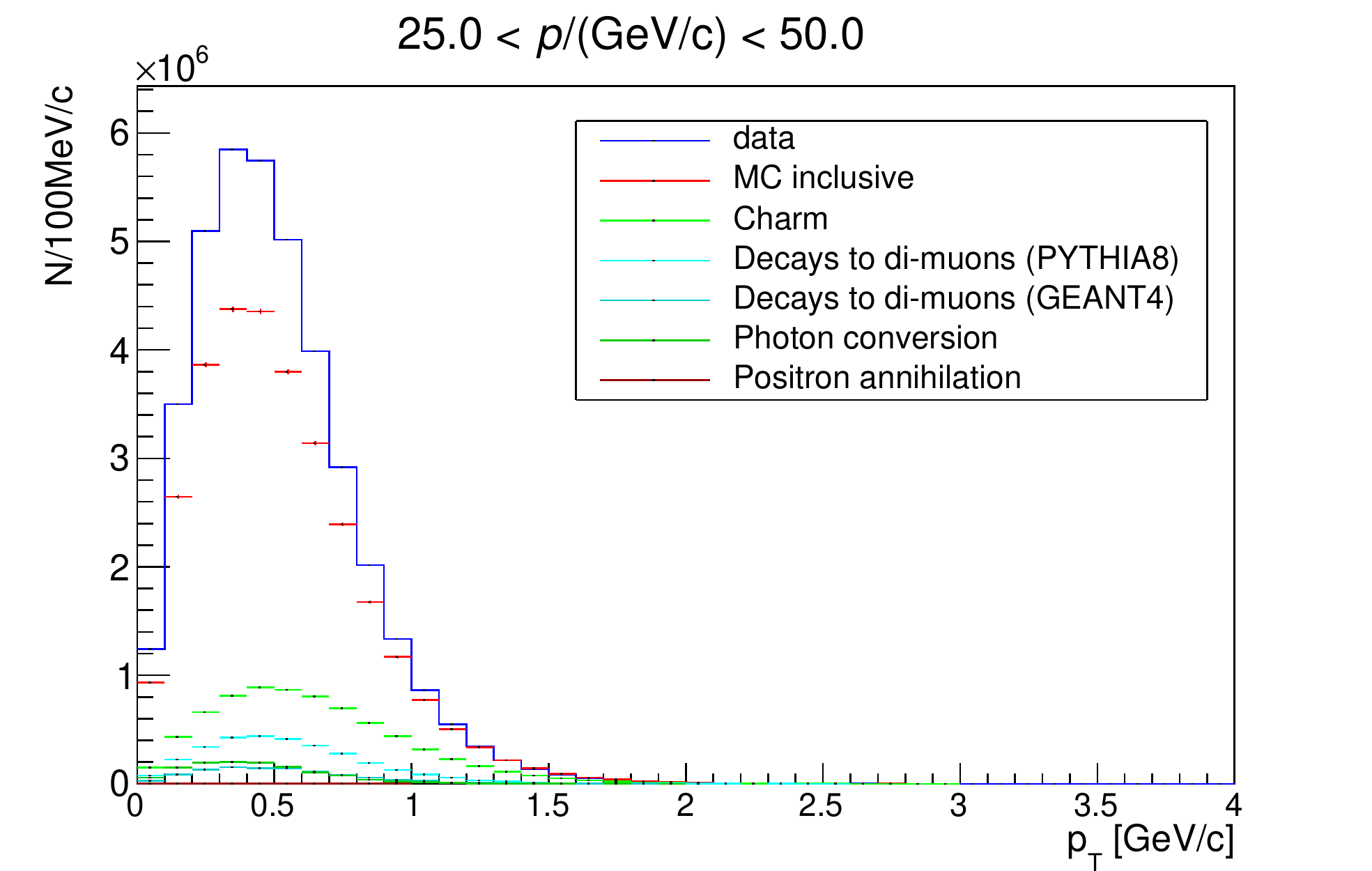}\includegraphics[width=0.4\textwidth]{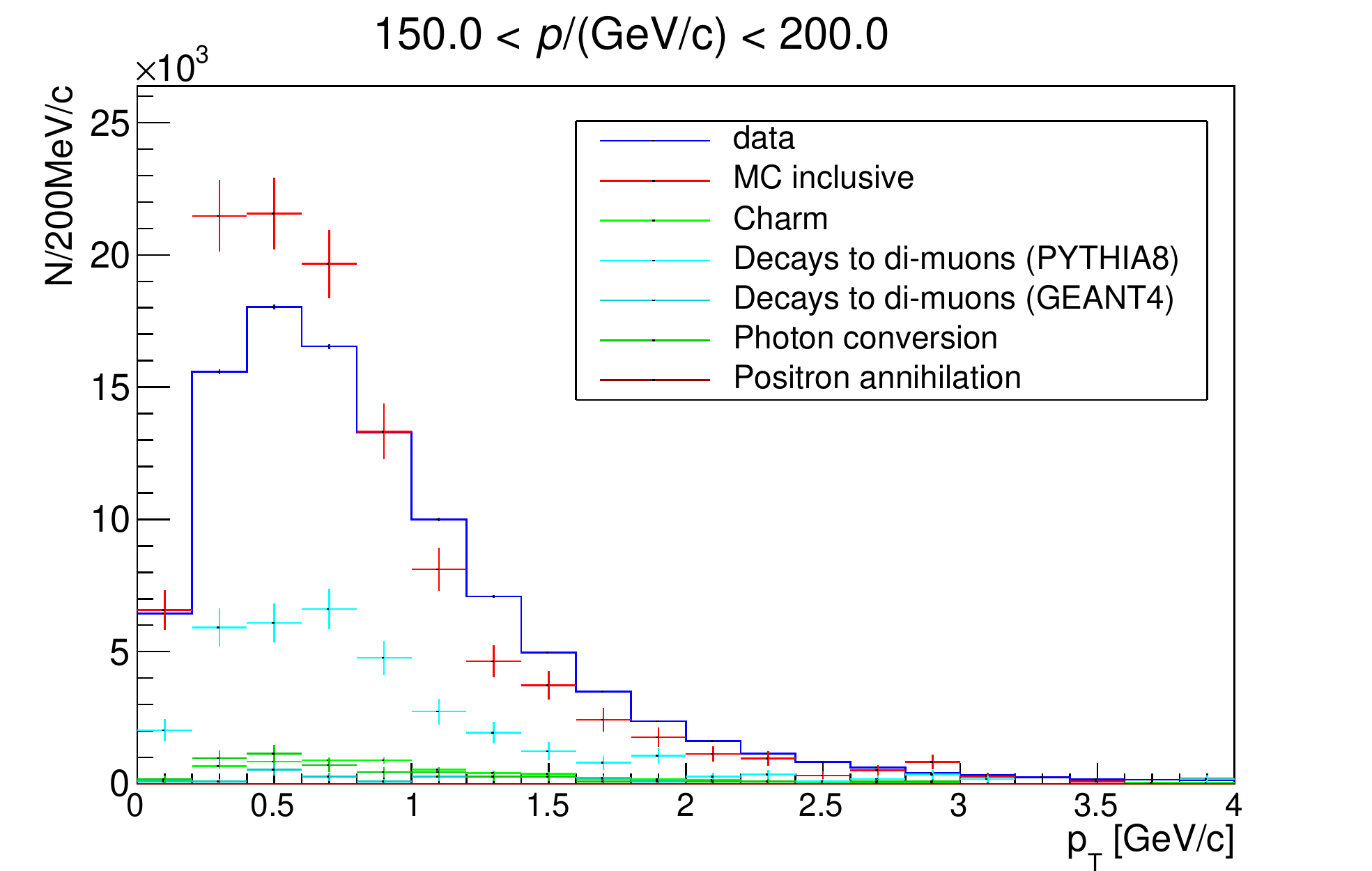}

\includegraphics[width=0.4\textwidth]{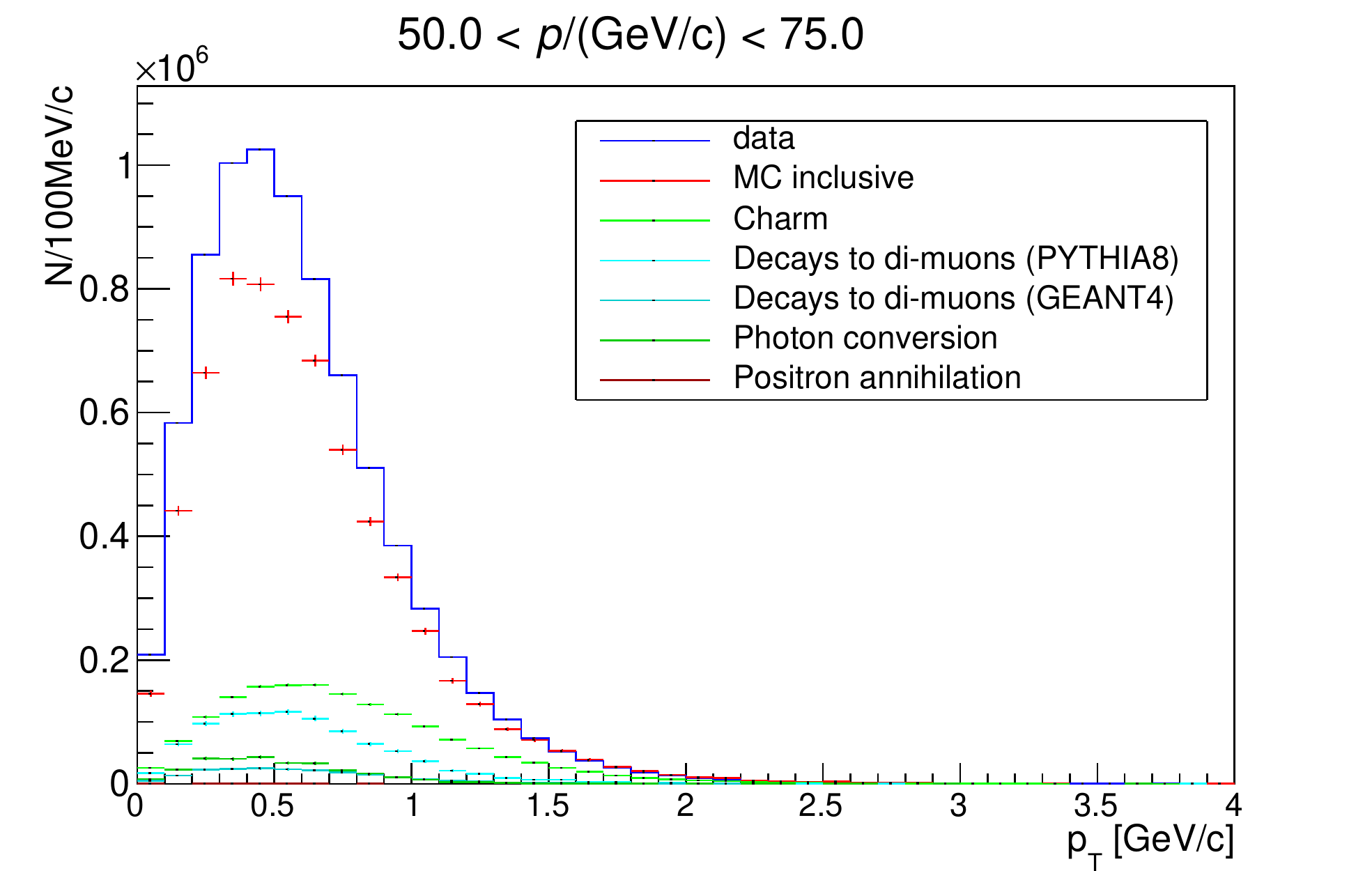}\includegraphics[width=0.4\textwidth]{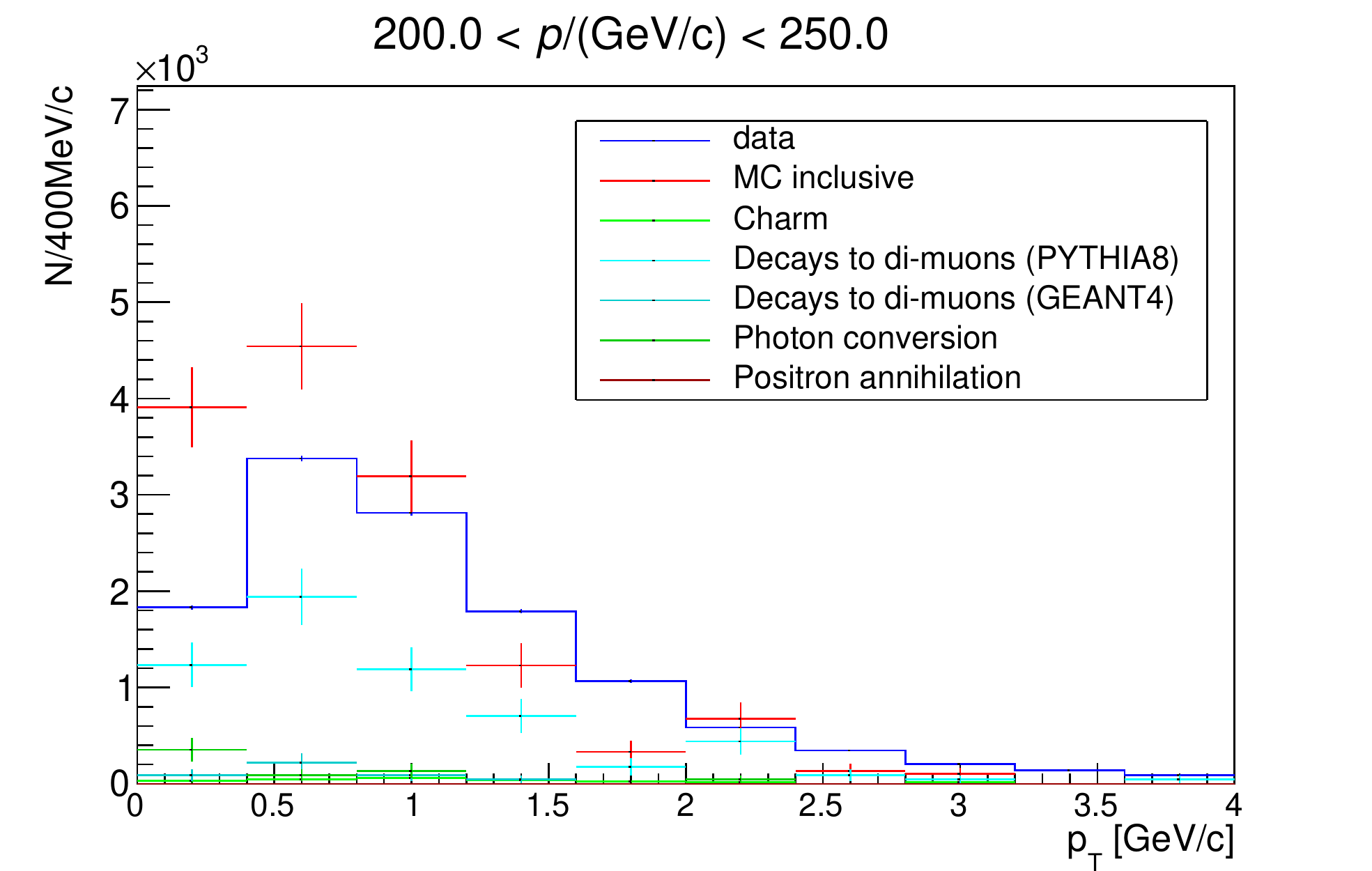}

\includegraphics[width=0.4\textwidth]{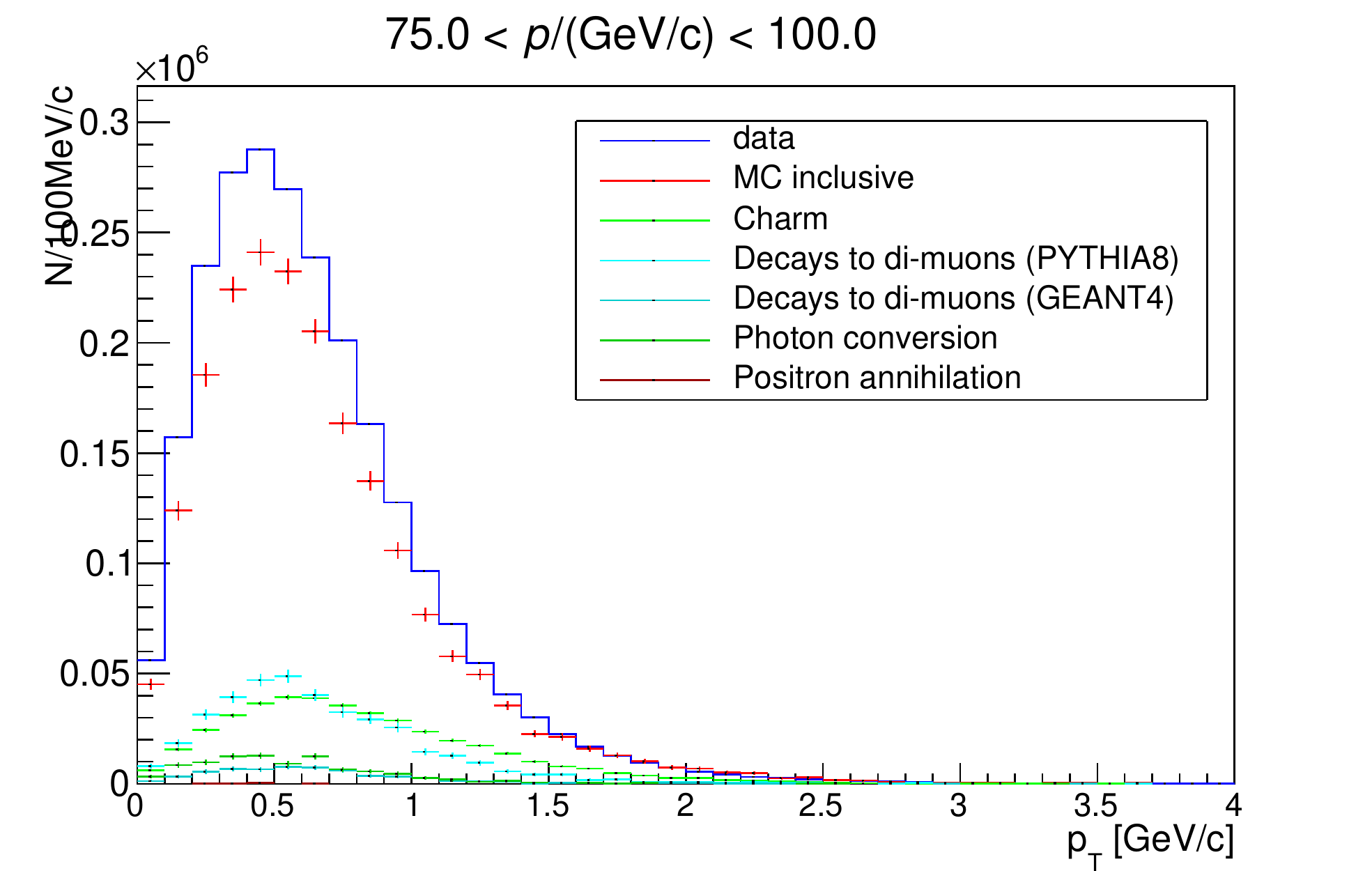}\includegraphics[width=0.4\textwidth]{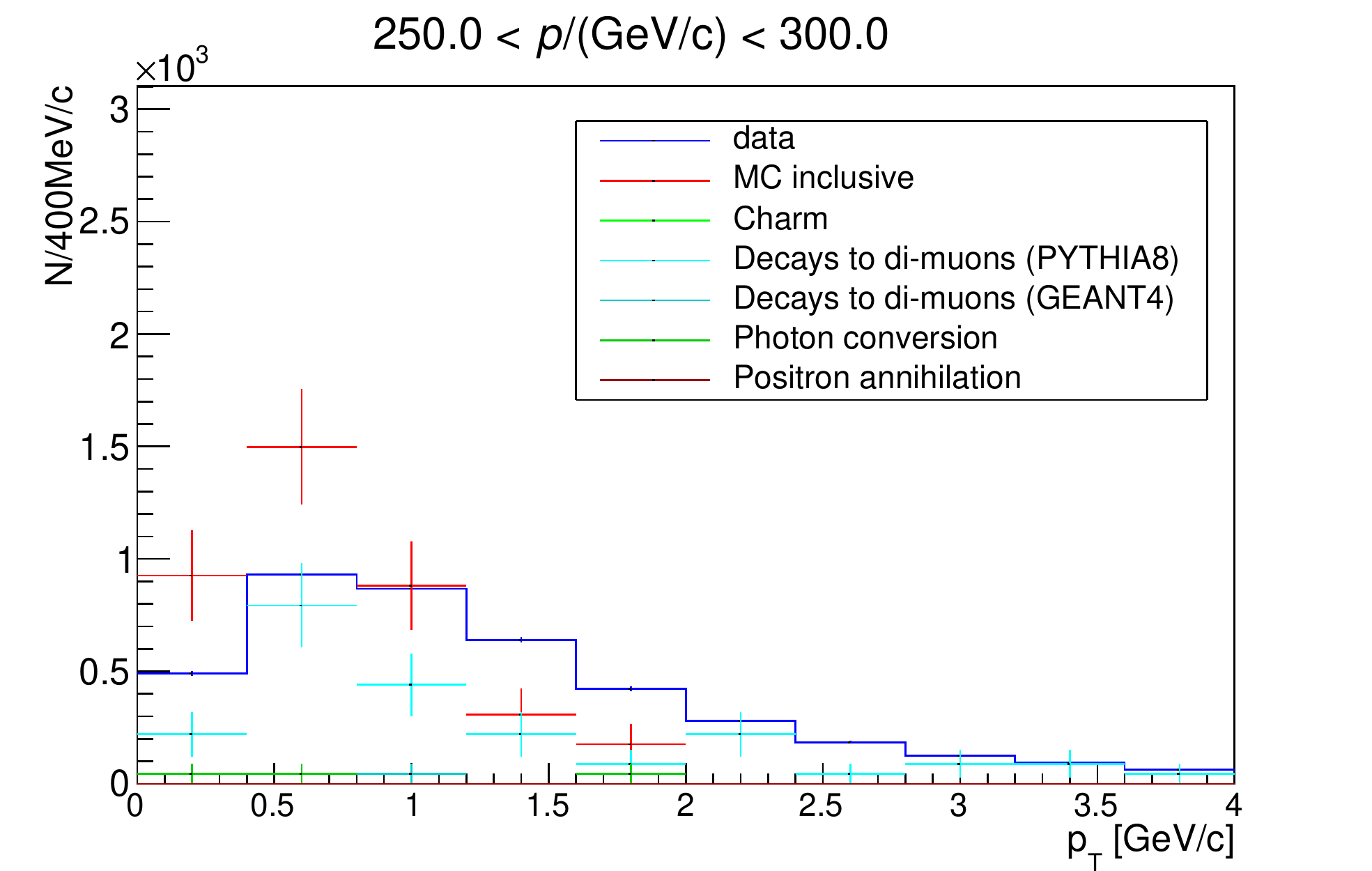}

\caption{$p_T$ distributions in slices of $p$ for data and simulation. The units on the vertical axes are the number of tracks per bin, with the simulation normalised to the data.}
\label{fig:ptslices}
\end{figure}

In Figure~\ref{fig:ptslices}, we show the $p_T$ distributions in slices of $p$. Table~\ref{tab:momBins} shows a numerical comparison of the number of tracks in the different momentum bins.

\begin{table}[ht]
\centering
\caption{Number of reconstructed tracks in different momentum bins per $10^9$ POT per \GeV/$c$ for data and simulation. The statistical errors for data are negligible. For data, the uncertainties are dominated by the uncertainty in the POT normalization, $2.1\%$. For the simulation, the main uncertainty is due to a different reconstruction efficiency in the simulation compared to data, $3.3\%$.}
\label{tab:momBins}
\begin{tabular}{|l| r | r | c |   }
\hline
Interval &data &Simulation & ratio  \\
\hline	
\hline	
$5-10$~\GeV/$c$     &$(1.13\pm0.02) \times 10^5$  & $(1.12\pm0.03)\times 10^5$  & $1.01 \pm  0.04$    \\
$10-25$~\GeV/$c$    &$(2.40\pm0.05) \times 10^4$  & $(1.85\pm0.06)\times 10^4$  & $1.29 \pm  0.05$    \\
$25-50$~\GeV/$c$    &$(4.80\pm0.10) \times 10^3$  & $(3.76\pm0.11)\times 10^3$  & $1.28 \pm  0.05$   \\
$50-75$~\GeV/$c$    &$(9.83\pm0.2)  \times 10^2$ & $(8.0\pm0.2)  \times 10^2$  & $1.23 \pm  0.05$   \\
$75-100$~\GeV/$c$   &$(2.95\pm0.06) \times 10^2$  & $(2.5\pm0.08) \times 10^2$  & $1.20 \pm  0.05$    \\
$100-125$~\GeV/$c$  &$(1.1\pm0.02)  \times 10^2$  & $(0.9\pm0.03) \times 10^2$  & $1.14 \pm  0.05$   \\
$125-150$~\GeV/$c$  &$21.0\pm0.4$                 & $20.1\pm7.5$                & $1.04 \pm  0.04$   \\
$150-200$~\GeV/$c$  &$6.4\pm0.1$                  & $6.6\pm0.3$                 & $0.96 \pm  0.04$   \\
$200-250$~\GeV/$c$  &$0.76\pm0.02$                & $0.88\pm0.06$               & $0.86 \pm  0.06$   \\
$250-300$~\GeV/$c$  &$0.26\pm0.01$                & $0.26\pm0.03$               & $0.97 \pm  0.11$   \\
\hline
\end{tabular}
\end{table}

\begin{figure}[ht]
\centering
\includegraphics[width=12.5cm]{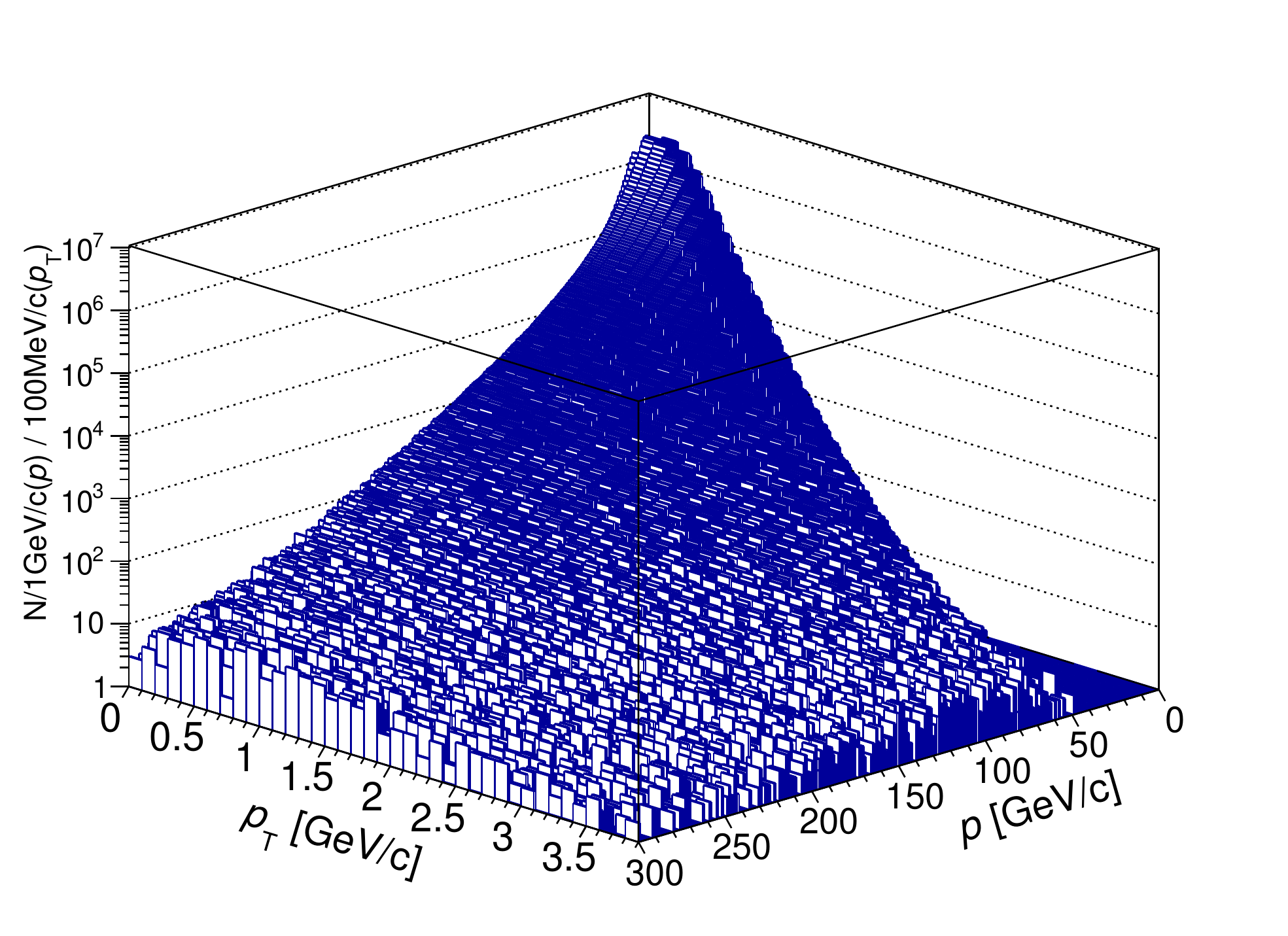}
\caption{$p_T$ vs $p$ for data. The units on the vertical axis are the number of tracks per $p,p_{T}$ bin in the entire data set.}
\label{fig:phasespace}
\end{figure}

Figure~\ref{fig:phasespace} shows the muon $p-p_{T}$ distribution in data.

\begin{figure}[ht]
\centering
\noindent\makebox[\textwidth]{
\includegraphics[width=15cm]{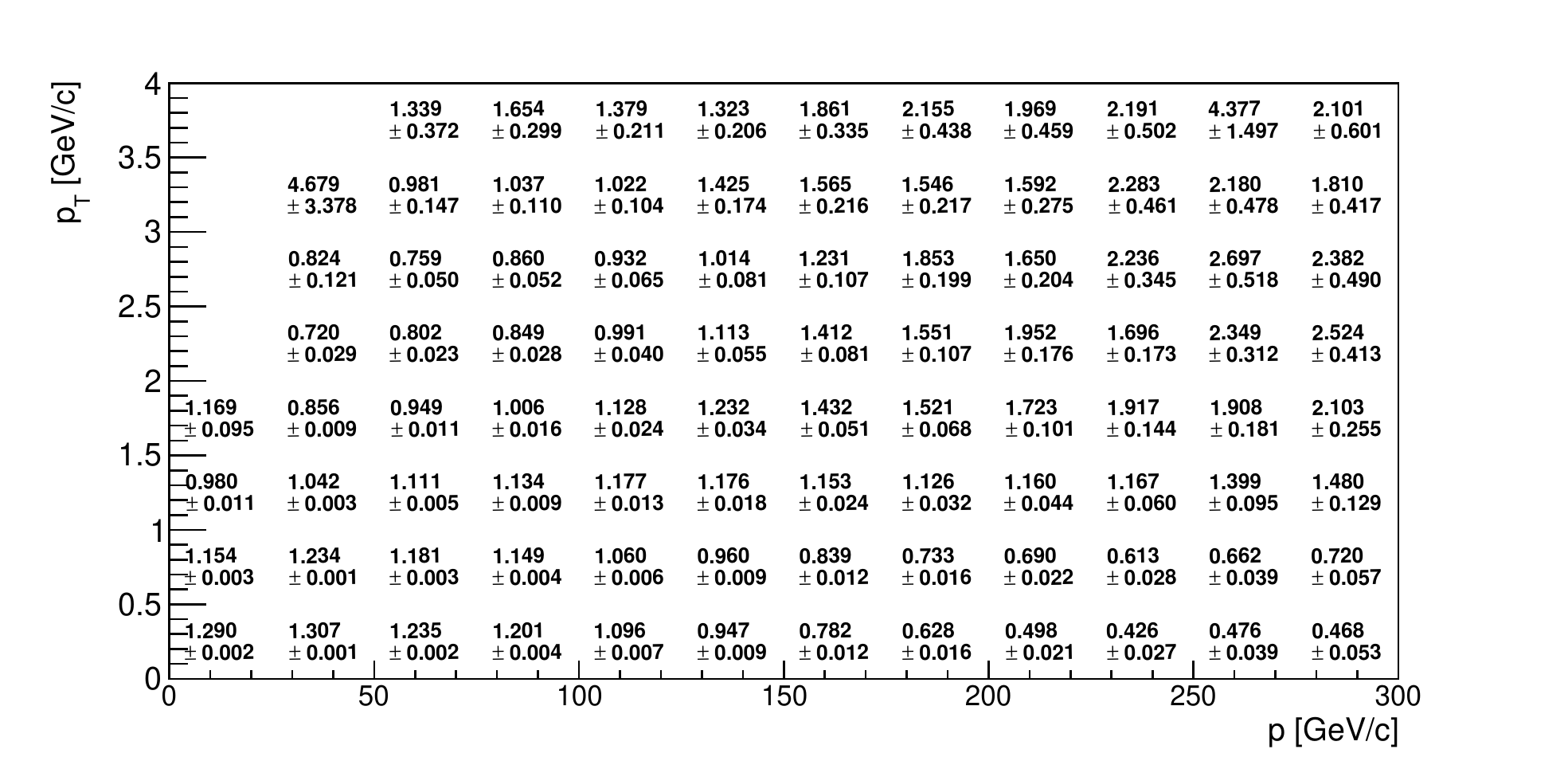}
}
\caption{Ratio of data and MC tracks, $R=\frac{N_{data}}{N_{MC}}$ in bins of $p$ and $p_T$.}
\label{fig:MCcomp2D}
\end{figure}

Figure~\ref{fig:MCcomp2D} gives a view of the differences between data and simulation in the $p-p_T$ plane. Plotted is the difference between number of data and simulated tracks divided by the sum of the tracks in data and simulation in bins of $p$ and $p_T$.

For momenta above $150$~\GeV/$c$, the simulation underestimates tracks with larger $p_T$, while the total number of tracks predicted are in agreement within $20\%$.
The difference between data and simulation is probably caused by a different amount of muons from pion and kaon decays. It was seen that by increasing the contribution of muons from pion and kaon decays in the simulation the difference between data and simulation was reduced.

The FLUKA~\cite{fluka1,fluka2} generator is used to determine the radiation levels in the SHiP environment. To validate the results from FLUKA, the muon flux setup was implemented in FLUKA and the simulation with this setup was compared to that made with Pythia/GEANT4. The results of this comparison are given in Annex~\ref{flukavsgeant}. This independent prediction provides additional support for the validity of the SHiP background simulation.  

\section{Conclusions}
\label{sec:summary}
We have measured the muon flux from  $400$~\GeV/$c$ protons impinging on a heavy tungsten/molybdenum target. The physics processes underlying this are a combination of the production of muons through decays of non-interacting pions and kaons, the production and decays of charm particles and low-mass resonances, and the transportation of the muons through $2.4~$m iron. Some 20--30\% differences in the absolute rates are observed. The simulation underestimates contributions to larger transverse momentum for higher muon momenta.  Given the complexity of the underlying processes, the agreement between the prediction by the simulation and the measured rate is remarkable.

Systematic errors for the track reconstruction ($3\%$) and POT normalization ($15~\rm{POT})/\mu\text{-}\rm{event}$ have been studied and estimated.

A further understanding of the simulation and the data will be obtained with an analysis of di-muon events, the results of which will be the subject of a future publication.

\section{Acknowledgments}
The SHiP Collaboration acknowledges support from the following Funding Agencies: the National Research Foundation of Korea
(with grant numbers of 2018R1A2B2007757, 2018R1D1A3B07050649, 2018R1D1A1B07050701, 2017R1D1A1B03036042,
2017R1A6A3A01075752, 2016R1A2B4012302, and 2016R1A6A3A11930680);  the Russian Foundation for Basic Research (RFBR, grant 17-02-00607) and the TAEK of Turkey.

This work is supported by a Marie Sklodowska-Curie Innovative Training Network Fellowship of the European Commissions Horizon 2020 Programme under contract number 765710 INSIGHTS.

We thank M. Al-Turany, F. Uhlig. S. Neubert and A. Gheata their assistance with FairRoot. We acknowledge
G. Eulisse and P.A. Munkes for help with Alibuild.

The measurements reported in this paper would not have been possible
without a significant financial contribution from CERN. In addition, several member institutes
made large financial and in-kind contributions to the construction of the target and the
spectrometer sub detectors, as well as providing expert manpower for commissioning, data
taking and analysis. This help is gratefully acknowledged.

\appendix
\section{FLUKA-GEANT4 comparison}
\label{flukavsgeant}
\subsection{Simulation samples}
The geometry of the muon flux spectrometer was reproduced in FLUKA with a few approximations \cite{flukacomp}.
A large sample of muons was generated for the comparison with GEANT4. For performance reasons three samples were made with different momentum thresholds (set for all particles). This increased the statistics in the corresponding momentum bins.
The number of POT for the three samples is shown in Table~\ref{tab:MCPOTFluka}.

\begin{table}[h]
\centering
\caption{FLUKA samples produced for Muon Flux comparison with GEANT4.}
\label{tab:MCPOTFluka}
\begin{tabular}{|l|l|l|} 
\hline
momentum threshold &POT&    Muon momentum range     \\
for transport of all particles &&\\
\hline	
5~\GeV/$c$ & $1.37~\times~10^8$& $5<p<30$ \GeV/$c$\\
27~\GeV/$c$ & $5.43~\times~10^8$& $30<p<100$ \GeV/$c$\\ 
97~\GeV/$c$ & $5.03~\times~10^8$& $p>100$ \GeV/$c$\\ 

\hline
\end{tabular}
\end{table}

The comparison is limited to 5~\GeV/$c<p<300$\GeV/$c$ and $p_T<4$~\GeV/$c$ to be consistent with  the GEANT4 simulations done for SHiP. \\
The primary proton-nuclei interactions are simulated and transported through the target and hadron absorber by FLUKA. Special settings of FLUKA were used to include:
\begin{itemize}
    \item full simulation of muon nuclear interactions and production of secondary hadrons;
     \item delta ray production from muons ($>$10 MeV);
     \item pair production and bremsstrahlung by high-energy muons;
     \item full transport and decay of charmed hadrons and tau leptons;
     \item decays of pions, kaons and muons described with maximum accuracy and polarisation.
 \end{itemize}
\subsection{Results}
In this section, we compare the reconstructed momentum distributions, $p$ and $p_T$, between FLUKA and GEANT4.

Tracks are considered to be muons if they have hits in the T1, T2, T3 and T4 stations. The distributions are taken at the T1 station and normalized to the number of POT. 

As shown in Figure~\ref{fig:momplotsP}, FLUKA predicts a lower rate compared to GEANT4. In the momentum range 5~\GeV/$c<p<200$~\GeV/$c$, the agreement between the two simulations is at the level of $\sim20\%$, above 200~\GeV/$c$ there is a discrepancy of a factor $\sim3$.

As shown in Figure~\ref{fig:momplotsPt}, FLUKA predicts a lower rate compared to GEANT4. In the transverse momentum range $0<p_T<1$ \GeV/$c$  the agreement between the two simulations is at the level of $\sim20\%$, while above 1~\GeV/$c$, there is a discrepancy of a factor $\sim3$.

\begin{figure}[ht]
\centering
\includegraphics[width=0.6\textwidth]{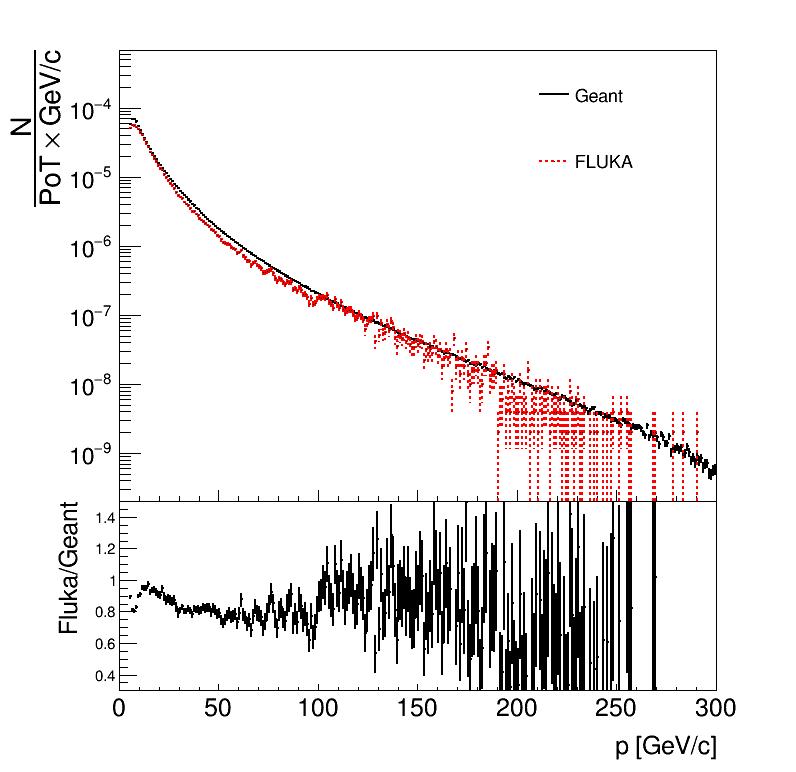}
\caption{Momentum distributions from FLUKA and GEANT4. The distributions are normalized to the number of POT.}
\label{fig:FlukamomplotsP}
\end{figure}

\begin{figure}[ht]
\centering
\includegraphics[width=0.6\textwidth]{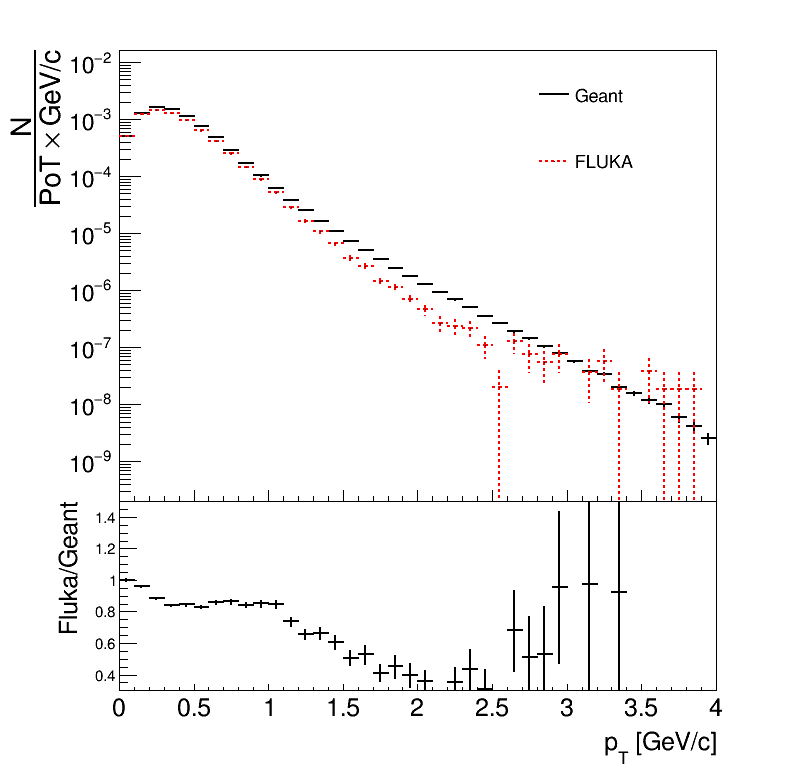}
\caption{Transverse momentum distributions from FLUKA and GEANT4. The distributions are normalized to the number of POT.  }
\label{fig:FlukamomplotsPt}
\end{figure}

Given the complexity of the processes underlying the production of muons and the approximations included in the geometry implementations, the agreement between the FLUKA and GEANT4 simulations is reasonable.
The differences between FLUKA and GEANT over the full muon momentum and transverse momentum spectra are within a factor 3. 
Therefore a safety factor of 3 is recommended for future radiological estimates related to muons in the SHiP environment.

\end{document}